\documentclass[useAMS,twocolumn,usenatbib,a4paper]{mn2e}

\def\OutputDriver{pdftex}
\usepackage[\OutputDriver]{graphicx}

\usepackage{amssymb}
\usepackage{amsmath}
\usepackage{bm}

\usepackage[\OutputDriver,hyperindex,bookmarks=false]{hyperref}
\hypersetup{
breaklinks  = {false},
colorlinks  = {false},
linkcolor   = {black},
pdfpagemode = {None},
pdfborder   = {0 0 1},
pdftitle    = {Abundances, masses, and weak-lensing mass profiles of galaxy clusters as a function of richness and luminosity in $\Lambda$CDM cosmologies},
pdfsubject  = {},
pdfauthor   = {S. Hilbert \& S.D.M. White},
pdfkeywords = {galaxies: general, galaxies: clusters: general, large-scale structure of the Universe, gravitational lensing, dark matter, cosmology: theory, methods: numerical}
}

\def\aj{AJ}%
\def\apj{ApJ}%
\def\apjl{ApJ}%
\def\apjs{ApJS}%
\def\aap{A\&A}%
\def\mnras{MNRAS}%
\def\nat{Nature}%
\def\physrep{Phys.~Rep.}%
\newcommand{\diff}[2][]{\mathrm{d}^{#1}{#2}}
\newcommand{\EV}[1]{\left\langle{#1}\right\rangle}
\newcommand{\ev}[1]{\langle{#1}\rangle}
\newcommand{\pdf}{\mathrm{pdf}}

\newcommand{\kpc}{\ensuremath{\mathrm{kpc}}}
\newcommand{\Mpc}{\ensuremath{\mathrm{Mpc}}}

\newcommand{\Msolar}{\ensuremath{\mathrm{M}_\odot}}
\newcommand{\Lsolar}{\ensuremath{\mathrm{L}_\odot}}

\newcommand{\degt}{\ensuremath{\mathrm{deg}}}

\newcommand{\Rcritmean}{\ensuremath{R^\text{crit/mean}_{200}}}

\newcommand{\Mcrit}{\ensuremath{M^\text{crit}_{200}}}
\newcommand{\Mmean}{\ensuremath{M^\text{mean}_{200}}}
\newcommand{\Mcritmean}{\ensuremath{M^\text{crit/mean}_{200}}}

\newcommand{\Nint}{\ensuremath{N^\text{gal}_\text{int}}} 
\newcommand{\Ngal}{\ensuremath{N^\text{gal}_{1\Mpc}}}
\newcommand{\Rgal}{\ensuremath{R^\text{gal}_{200}}}
\def\N200{N^\text{gal}_{200}}
\def\L200{L^\text{gal}_{200}}

\begin{document}

\title[Galaxy cluster properties in $\Lambda$CDM cosmologies]{Abundances, masses, and weak-lensing mass profiles of galaxy clusters as a function of richness and luminosity in $\Lambda$CDM cosmologies}
\author[S. Hilbert \& S.D.M. White]{
Stefan Hilbert$^{1,2}$\thanks{\texttt{shilbert@astro.uni-bonn.de}} and
Simon D.~M. White$^{2}$
\\$^{1}$Argelander-Institut f{\"u}r Astronomie, Auf dem H{\"u}gel 71, 53121 Bonn, Germany
\\$^{2}$Max-Planck-Institut f{\"u}r Astrophysik,Karl-Schwarzschild-Stra{\ss}e 1, 85741 Garching, Germany
}
\date{\today}
\maketitle

\begin{abstract}
%Aims:
We test the concordance $\Lambda$CDM cosmology by comparing predictions for the mean properties of galaxy clusters to observations. 
%Methods:
We use high-resolution $N$-body simulations of cosmic structure formation and semi-analytic models (SAMs) of galaxy formation to compute the abundance, mean density profile, and mass of galaxy clusters as a function of richness and luminosity, and we compare these predictions to observations of clusters in the Sloan Digital Sky Survey (SDSS) maxBCG catalogue. We discuss the scatter in the mass-richness relation, the reconstruction of the cluster mass function from the mass-richness relation, and fits to the weak-lensing cluster mass profiles. The impact of cosmological parameters on the predictions is investigated by comparing results from galaxy models based on the Millennium Simulation (MS) and another WMAP1 simulation to those from a WMAP3 simulation. 
%Results:
We find that the simulated weak-lensing mass profiles and the observed profiles of the SDSS maxBCG clusters agree well in shape and amplitude. The mass-richness relations in the simulations are close to the observed relation, with differences $\lesssim 30\%$. The MS and WMAP1 simulations yield cluster abundances similar to those observed, whereas abundances in the WMAP3 simulation are 2-3 times lower. 
%Conclusions:
The differences in cluster abundance, mass, and density amplitude between the simulations and the observations can be attributed to differences in the underlying cosmological parameters, in particular the power spectrum normalisation $\sigma_8$. Better agreement between predictions and observations should be reached with a normalisation $0.722<\sigma_8<0.9$ (probably closer to the upper value), i.e. between the values underlying the two simulation sets. 
\end{abstract}


%%%%%%%%%%%%%%%%%%%%%%%%%%%%%%%%%%%%%%%%%%%%%%%%%
\section{Introduction}
\label{sec:Introduction}
%%%%%%%%%%%%%%%%%%%%%%%%%%%%%%%%%%%%%%%%%%%%%%%%%

Clusters of galaxies are a powerful probe of astrophysics and cosmology \citep[][]{Voit2005}. For example, the cluster mass function is very sensitive to the cosmic mean matter density, to the initial fluctuation amplitude \citep[][]{PressSchechter1974,FrenkEtal1990,EkeColeFrenk1996,ShethTormen1999} and to dark energy dynamics \citep[][]{BartelmannDoranWetterich2006,GrossiSpringel2009,FrancisLewisLinder2009}. It can be predicted to high accuracy by numerical simulations \citep[][]{JenkinsEtal2001,WarrenEtal2006,LukicEtal2007}.

However, the masses of galaxy clusters cannot be observed directly. Thus one either needs to infer the masses of observed clusters from some more directly observable cluster property, requiring accurate knowledge of the observable-mass relation and its scatter, or one must directly compare observed and predicted cluster abundance as a function of such observables. These include the X-ray luminosity and temperature of the intracluster gas \citep[][]{BorganiEtal2001,ReiprichBoehringer2002,StanekEtal2006,PiffarettiValdarnini2008,VikhlininEtal2009}, the number and velocity dispersion of the cluster galaxies \citep[][]{Zwicky1937_gal_and_cluster_masses,RinesEtal2003,KochanekEtal2003,BeckerEtal2007}, the number of giant arcs \citep[][]{BartelmannEtal1998,WambsganssBodeOstriker2004,FedeliEtal2008}, the weak-lensing signal \citep[][]{TysonWenkValdes1990,CyprianoEtal2004,Hoekstra2007,JohnstonEtal2007_SDSS_cluster_wl_II_arXiv,ReyesEtal2008}, and the Sunyaev-Zel'dovich signal induced by the cluster \citep[][]{WhiteHernquistSpringel2002,SchulzWhite2003,BonaldiEtal2007,StaniszewskiEtal2009}.

Predictions for many cluster observables (e.g. the X-ray luminosity or the cluster richness) and for their relation to the cluster mass require modelling of astrophysical processes such as gas cooling and galaxy formation. Although inaccuracies in such astrophysical models are an unpleasant source of uncertainty for cosmological parameter estimation, they mean that comprehensive observations of clusters can be used to constrain cosmological parameters and models for cluster/galaxy evolution simultaneously.

The largest sample of observed galaxy clusters currently available is the maxBCG cluster catalogue \citep{KoesterEtal2007_SDSS_clusters}, This was extracted from the Sloan Digital Sky Survey (SDSS)\footnote{\texttt{http://www.sdss.org}} using maxBCG, an optical cluster-finding algorithm \citep{KoesterEtal2007_MaxBCG}. Constraints on the scatter in the velocity dispersion-richness relation and the mass-richness relation of these clusters have been derived from cluster X-ray and galaxy velocity dispersion observations \citep{BeckerEtal2007,RozoEtal2009_MaxBCG_III_scatter}. Weak-lensing measurements of average cluster mass profiles have also been used to calibrate the mass-richness relation, the mass-optical luminosity relation, and mass-to-light ratio profiles of the maxBCG clusters \citep{SheldonEtal2009_SDSS_cluster_wl_I,SheldonEtal2009_SDSS_cluster_wl_III,JohnstonEtal2007_SDSS_cluster_wl_II_arXiv,ReyesEtal2008}. These data provide significant constraints on cosmological parameters \citep{RozoEtal2009_MaxBCG_V_cosmology_arXiv}.

In this work, we investigate how well physically based models for galaxy formation in a $\Lambda$CDM universe can reproduce the observed relations of cluster richness and luminosity to other cluster properties, most notably mass. We also investigate what information on cosmological parameters and galaxy evolution can be obtained by comparing model predictions to observation. We use the Millennium Simulation \citep[][]{SpringelEtal2005_Millennium} and two smaller $N$-body simulations of cosmic structure formation \citep[][]{WangEtal2008} in conjunction with semi-analytic models of galaxy evolution \citep[][]{DeLuciaBlaizot2007,WangEtal2008} to create mock catalogues of galaxy clusters selected similarly to the maxBCG catalogue. We compute cluster abundances, average cluster masses, and weak-lensing mass profiles as a function of cluster richness and luminosity, and we compare these to observational results for the SDSS maxBCG sample. In addition, we investigate the scatter in the mass-richness relation, and we discuss how well one can recover the cluster mass function and the weak-lensing mass profiles from the richness-binned cluster abundances and mean masses.

The semi-analytic galaxy models used here couple star formation in galaxies to the properties of the evolving dark matter halo distribution in which the galaxies live. The models have been adjusted to be consistent with various observations, e.g. the luminosities, stellar masses, morphologies, gas contents and correlations of galaxies at low redshift, but they have not been tuned to match the properties of rich clusters.
The comparison to observations provided here is thus a direct test of these models and their description of the physical processes relevant for galaxy formation. This contrasts with halo occupation distribution models \citep{CooraySheth2002}, where the galaxy populations of clusters are adjusted to fit observation without considering in detail how they could have been built up by physical processes within the evolving dark matter distribution.

Our paper is organised as follows. We discuss the $N$-body simulations and galaxy models, as well as our methods for creating the simulated cluster samples from them in Sec.~\ref{sec:methods}. Results for our simulated cluster samples and a comparison to observation are presented in Sec.~\ref{sec:results}. Our paper concludes with a summary and discussion in Sec.~\ref{sec:discussion}.

%%%%%%%%%%%%%%%%%%%%%%%%%%%%%%%%%%%%%%%%%%%%%%%%%
\section{Methods}
\label{sec:methods}
%%%%%%%%%%%%%%%%%%%%%%%%%%%%%%%%%%%%%%%%%%%%%%%%%

We use cosmological $N$-body simulations to analyse the matter distribution in and around galaxy clusters in two different $\Lambda$CDM cosmologies. We infer the properties of the galaxies in the clusters from model galaxy catalogues generated by applying semi-analytic galaxy formation models to halo assembly trees generated from the outputs of the $N$-body simulations. We then compute richness and luminosity estimates for clusters in the model galaxy catalogues taking into account several observational features of optical cluster-finding algorithms, in particular maxBCG by \citet{KoesterEtal2007_MaxBCG}.

%%%%%%%%%%%%%%%%%%%%%%%%%%%%%%%%%%%%%%%%%%%%%%%%%
\subsection{The $N$-body simulations}
\label{sec:N_body_simulations}
%%%%%%%%%%%%%%%%%%%%%%%%%%%%%%%%%%%%%%%%%%%%%%%%%

%================================================
\begin{table}
\center
\caption{
\label{tab:simulation_cosmological_parameters}
The cosmological parameters at redshift 0 for the three simulations used in this study. The parameters are: the baryon density $\Omega_\mathrm{b}$, the matter density $\Omega_\mathrm{M}$, and the energy density of the cosmological constant $\Omega_\Lambda$ (in units of the critical density), the Hubble constant $h$ (in units of $100\,\mathrm{km}\mathrm{s}^{-1}\Mpc^{-1}$), the primordial spectral index $n$ and the normalisation parameter $\sigma_8$ for the linear density power spectrum.
} 
\begin{tabular}{l l l l}
\hline
\hline
                          & MS \& WMAP1 & WMAP3 \\
\hline
$\Omega_\mathrm{b}$       & 0.045       & 0.04  \\
$\Omega_\mathrm{M}$       & 0.25        & 0.226 \\
$\Omega_\mathrm{\Lambda}$ & 0.75        & 0.774 \\
$h$                       & 0.73        & 0.743 \\
$n$                       & 1           & 0.947 \\
$\sigma_8$                & 0.9         & 0.722 \\
\hline
\end{tabular}
\end{table}
%================================================

Our study is based on three different $N$-body simulations: the Millennium Simulation (MS) by \citet{SpringelEtal2005_Millennium}, and two smaller simulations WMAP1 and WMAP3 by \citet{WangEtal2008}.\footnote{We refer the reader to \citet{SpringelEtal2005_Millennium}, \citet{WangEtal2008}, and references therein for more details about the simulations.}
The simulations assume a flat $\Lambda$CDM cosmology with parameters given in Table~\ref{tab:simulation_cosmological_parameters}. Both the MS and the WMAP1 simulation use a parameter set that was derived by combining the WMAP 1st-year results \citep{SpergelEtal2003_WMAP_1stYear_Data} with results from the 2dFGRS \citep{CollessEtal2003_2dF_Data}. The WMAP3 simulation employs cosmological parameters that are consistent with data from the WMAP 3rd-year release, the Cosmic Background Imager, and the Very Small Array \citep{SpergelEtal2007_WMAP_3rdYear_Data}, with a bias towards values differing from those used for the other two simulations.

The most prominent differences between the two sets of cosmological parameters are the normalisation parameter $\sigma_8$ and the spectral index $n$ for the density power spectrum: The MS and the WMAP1 simulation assume $\sigma_8=0.9$ and $n=1$, whereas the WMAP3 simulation assumes lower values $\sigma_8=0.722$ and $n=0.94$. Hence, there is less power on small scales in the matter power spectrum of the WMAP3 simulation than in the MS and the WMAP1 simulation. This results in a substantial delay in structure formation and less massive collapsed structures at any given redshift in the WMAP3 simulation.

%================================================
\begin{table}
\center
\caption{
\label{tab:simulation_numericial_parameters}
Numerical parameters for the three simulations used in this study. The parameters are: the comoving cube size $L$, the particle number $n_\mathrm{p}$, the particle mass $m_\mathrm{p}$, and the effective force softening length $\epsilon$.
}
\begin{tabular}{l l l l}
\hline
\hline
                                 & MS              & WMAP1          & WMAP3            \\
\hline
$L$ [$h^{-1}\Mpc$]               & 500             & 125             & 125             \\
$n_\mathrm{p}$                   & $2160^3$        & $560^3$         & $560^3$         \\
$m_\mathrm{p}$ [$h^{-1}\Msolar$] & $8.6\times10^8$ & $8.6\times10^8$ & $7.8\times10^8$ \\
$\epsilon$ [$h^{-1}\kpc$]        & 5               & 5               & 5               \\
\hline
\end{tabular}
\end{table}
%================================================

The simulations were run using a parallel TreePM version of \textsc{GADGET2} \citep{Springel2005_GADGET2}. The numerical parameters of the simulations are listed in Table~\ref{tab:simulation_numericial_parameters}. The main difference between the MS and the WMAP simulations is the simulation box size. The large volume of the MS provides us with a large sample of galaxies and galaxy clusters with statistical errors comparable to the `cosmic variance' errors of the SDSS maxBCG sample. The smaller WMAP simulations differ in their cosmological parameters, but share the same numerical parameters and initial conditions.\footnote{
Their initial density fields are identical except for small amplitude adjustments needed to reproduce the correct matter power spectra.
}
This reduces the influence of sampling noise when comparing results between the WMAP simulations and allows us to study the influence of cosmology on our results.

For each simulation, the particle data were stored on disk at 64 output times. These snapshots contain information on dark-matter halos, which have been identified running a friend-of-friend (FOF) group finding algorithm on the set of simulation particles. The halos were later decomposed into subhalos using \textsc{SUBFIND} \citep{SpringelEtal2001_SUBFIND} to identify gravitationally self-bound locally overdense regions. The most massive subhalo, called the main subhalo, typically contains 90\% of the FOF halo mass and shares its centre. Detailed merging history trees of all self-bound dark-matter subhalos were then computed. The resulting merger trees were used as input for the semi-analytic models discussed in the following section.

%%%%%%%%%%%%%%%%%%%%%%%%%%%%%%%%%%%%%%%%%%%%%%%%%
\subsection{The semi-analytic galaxy models}
\label{sec:SAMs}
%%%%%%%%%%%%%%%%%%%%%%%%%%%%%%%%%%%%%%%%%%%%%%%%%

We use semi-analytic galaxy formation models from the Munich family \citep{KauffmannEtal1999_I,SpringelEtal2001_SUBFIND,DeLuciaKauffmannWhite2004, SpringelEtal2005_Millennium, CrotonEtal2006,DeLuciaBlaizot2007} to set the optical properties of galaxies in the $N$-body simulations. These semi-analytic models assume that galaxies form from gas that accumulates in the centre of each dark-matter subhalo in the simulation. Star formation in each galaxy is coupled to its subhalo properties via simple prescriptions for gas cooling, star formation, chemical enrichment and feedback from supernova and central black holes (AGNs), as well as for merging of galaxies once their dark matter subhalos have merged. Certain parameters quantifying the efficiency of these processes can be adjusted in order to maximise agreement of the results with observation.

For the MS, we use the publicly available galaxy catalogue\footnote{http://www.mpa-garching.mpg.de/Millennium} that was generated using the galaxy model described in \citet{DeLuciaBlaizot2007}. This MS model reproduces various observed relations for galaxies, in particular the observed luminosity, colour, gas content and morphology distributions \citep{CrotonEtal2006,DeLuciaEtal2006,KitzbichlerWhite2007} and the observed two-point correlation functions \citep{SpringelEtal2005_Millennium,KitzbichlerWhite2007}.

The galaxy model of \citet{DeLuciaBlaizot2007} has been applied by \citet{WangEtal2008} to the WMAP1 simulation using the same set of efficiency parameters. As expected, this gives similarly good agreement with observations as for the MS. Here, we will use these model galaxies, which we refer to below as the WMAP1-A model, for the WMAP1 simulation.

\citet{WangEtal2008} also applied the galaxy modelling technique of \citet{DeLuciaBlaizot2007} to the WMAP3 simulation, but with two slightly different sets of efficiency parameters. One, which we call WMAP3-B here, employs the same star formation efficiency as the WMAP1-A model, but lower supernova and AGN feedback efficiencies. The other model, called WMAP3-C in the following, employs a higher star-formation efficiency but also higher feedback efficiencies. Both the WMAP3-B and the WMAP3-C model show good agreement with low-redshift galaxy observations, e.g. of the galaxy luminosity function and the galaxy clustering \citep{WangEtal2008}. In the following, we will consider both models for the WMAP3 simulation.

%%%%%%%%%%%%%%%%%%%%%%%%%%%%%%%%%%%%%%%%%%%%%%%%%
\subsection{The ridgeline galaxies}
\label{sec:ridgeline_galaxies}
%%%%%%%%%%%%%%%%%%%%%%%%%%%%%%%%%%%%%%%%%%%%%%%%%

Optical cluster finding algorithms such as maxBCG \citep{KoesterEtal2007_MaxBCG} locate galaxy clusters by searching for local overdensities of E/S0 ridgeline galaxies in angular and redshift space. These galaxies are common in known clusters, and they are relatively easy to find, since many of them are bright and their observed colours are strongly correlated with luminosity and redshift. Because of their small scatter in colour, a narrow search range in colour can be chosen at each redshift. This narrow search range forces one to know the mean ridgeline colours rather accurately. Accurate mean ridgeline colours are also important for accurately quantifying the ridgeline galaxy content of the clusters.

Although the semi-analytic models can reproduce many of the observed properties of galaxies, there are still some discrepancies. In particular, the models do not reproduce the colours of passively evolving galaxies to the degree required for a direct application of the ridgeline colour-redshift relation used for the maxBCG catalogue \citep[see, e.g.,][for possible reasons]{WeinmannEtal2006}. We therefore `measure' the mean ridgeline colours of model galaxies as a function of redshift:
We roughly identify the mean colour of the ridgeline galaxies in the colour-magnitude diagram (where the ridgeline population induces a visible overdensity among the bright galaxies with colours close to the observed mean ridgeline colour). We then fit a Gaussian with mean $\bar{x}$ and variance $\sigma^2$ to the distribution of galaxy colours in a region around our initial colour guess.
We repeat the fit considering all galaxies with absolute $i$-band magnitudes $M_i\leq-20$ and colours in the range $\bar{x}\pm 3 \sigma$ until $\bar{x}$ converges.\footnote{
The dependence of the mean ridgeline colour on magnitude is ignored in this procedure, since for all models the slope in the colour-magnitude relation is very small in the considered colours and magnitude range.}

\begin{figure}
\centerline{\includegraphics[width=1\linewidth]{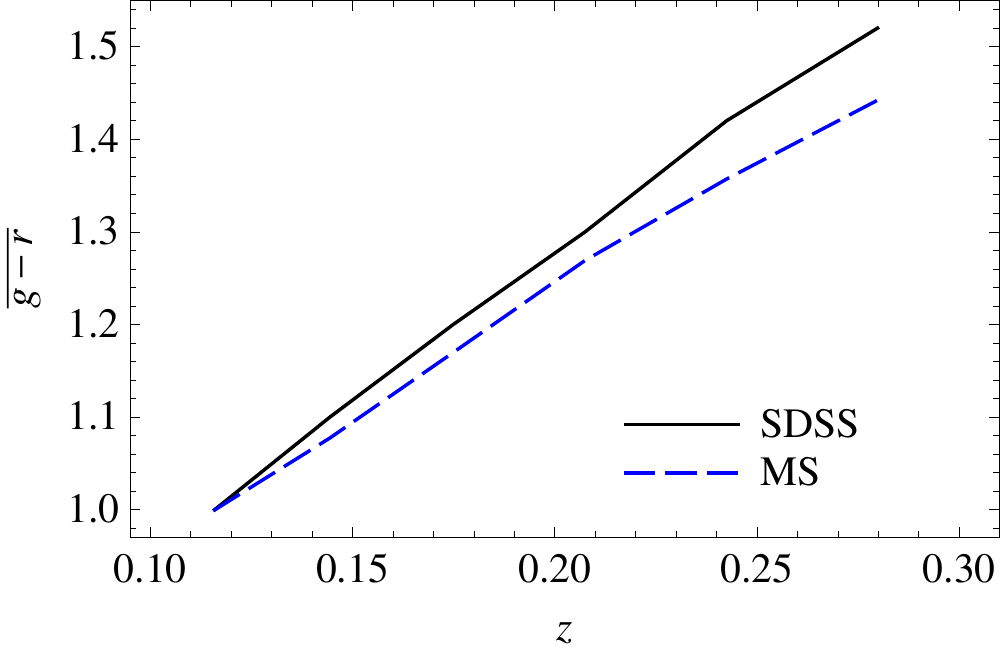}}
\caption{
\label{fig:ridgeline_colors}
The mean colour $\overline{g-r}(z)$ of the ridgeline galaxies as a function of redshift $z$ for the galaxy model of \protect\citet{DeLuciaBlaizot2007} in the MS compared to the mean colour-redshift relation for the ridgeline galaxies in the SDSS \protect\citep[][]{KoesterEtal2007_MaxBCG}.
}
\end{figure}

The resulting mean ridgeline colours for model galaxies in the MS as a function of redshift is compared to the corresponding relation for SDSS galaxies in Fig.~\ref{fig:ridgeline_colors}. The mean ridgeline colours of the model are close the observed ridgeline colours at low redshift, but they deviate significantly at higher redshift. The measured ridgeline width  $\sigma_{g-r}\approx0.05$ for all considered galaxy models and redshifts, which is in agreement with observations \citep{KoesterEtal2007_MaxBCG}. The measured width in $(r-i)$-colour $\sigma_{r-i}\approx0.03$, which is smaller than the observed value $\sigma_{r-i}=0.06$.

%================================================
%================================================

Following the maxBCG observational procedure, we consider $g-r$ and $r-i$ colours to identify ridgeline galaxies in our simulations. For each simulation snapshot with redshift $0.1\leq z \leq 0.3$, we select all objects in our semi-analytic galaxy catalogues whose $g-r$ and $r-i$ values are both within $2\sigma$ of the mean ridgeline colours. As mean colours, we take the values measured from the simulations. For the ridgeline width $\sigma$, we take the observed values $\sigma_{g-r}=0.05$ and $\sigma_{r-i}=0.06$ \citep{KoesterEtal2007_MaxBCG}.

Besides colour, galaxy brightness is used to select ridgeline galaxies. We thus further select from all model galaxies surviving the colour selection those with apparent observer-frame $i$-band magnitude $i^\text{obs}\leq i^\text{obs}_\text{lim}$. Here, we employ the same magnitude limit $i^\text{obs}_\text{lim}$ as \citet{KoesterEtal2007_MaxBCG} (B. Koester, private communication). This magnitude limit corresponds to an absolute rest-frame magnitude limit $M^\text{rest}_i \approx -20.25 + 5\log_{10}h \approx -20.9$ for the cosmologies considered here.

%%%%%%%%%%%%%%%%%%%%%%%%%%%%%%%%%%%%%%%%%%%%%%%%%
\subsection{The galaxy clusters}
\label{sec:model_clusters}
%%%%%%%%%%%%%%%%%%%%%%%%%%%%%%%%%%%%%%%%%%%%%%%%%

We identify galaxy clusters in the simulations with dark matter halos found by the FOF algorithm. From the simulation data stored on disk, we obtain for each such halo (hence cluster candidate) the positions of the centre, the virial radius $\Rcritmean$ (i.e. the radius of the largest sphere enclosing $200\times$ the critical/mean density) and the mass $\Mcritmean$ (i.e. the mass enclosed within $\Rcritmean$).

The semi-analytic galaxy models provide us with information about the galaxies associated with each dark matter halo.
For each halo, we measure the total number $\Nint$ of associated ridgeline galaxies selected as described in the preceding section. In addition, we count the number of ridgeline galaxies $\Ngal$ within a physical radius of $1h^{-1}\,\Mpc$ in projection along a simulation box axis. The result is used to compute the ``observationally defined'' radius $\Rgal = 0.156 (\Ngal)^{0.6} h^{-1}\,\Mpc$ \citep[see][]{HansenEtal2005,KoesterEtal2007_MaxBCG}. We then calculate a scaled galaxy richness $\N200$ by counting all ridgeline galaxies within a projected radius $\Rgal$. A cluster luminosity $\L200$ is then computed by summing the $i$-band luminosities of all ridgeline galaxies within $\Rgal$. These procedures mimic closely those employed to estimate richnesses and luminosities for the real maxBCG catalogue.

To improve statistics, we perform the measurements of the projected quantities $\Ngal$, $\Rgal$, $\N200$, and $\L200$ along all three simulation box axes. Each projection is treated as an individual cluster in the subsequent analysis. For the computation of cluster densities, the resulting triplication of clusters is taken into account by assuming a three times larger simulation volume.

Several effects hamper a direct comparison between observed clusters and our simulated clusters at the stage described so far. Here, we will not take into account fragmentation, but we do correct for contamination of the observational data by foreground and background galaxies.\footnote{
Fragmentation is insignificant in the maxBCG sample, but overmerging slightly boosts the cluster richness estimates mainly due to contamination by foreground and background structures \citep[][]{KoesterEtal2007_MaxBCG}.}
Using spectroscopic data, \citet{KoesterEtal2007_SDSS_clusters} found that roughly 16\% of the galaxies identified by maxBCG as cluster ridgeline galaxies are, in fact, projections. We include such projections in our simulated clusters in a very simple way, by randomly duplicating about 19\% of the ridgeline galaxies. As a result, our model clusters appear to be contaminated at the 16\% level.

%================================================
\begin{figure}
\centerline{\includegraphics[width=1\linewidth]{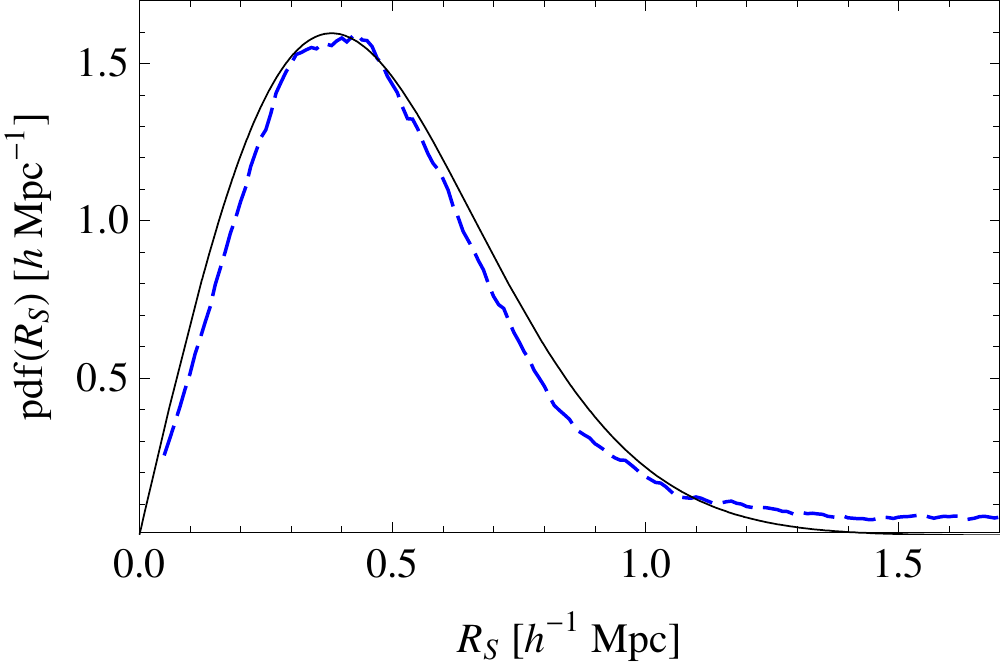}}
\caption{
\label{fig:offset_distribution}
The distribution of projected offset $R_s$ between the `true' and 'apparent' centres of miscentred clusters in the MS (dashed line). The distribution is well fit by a 2D Gaussian (solid line).
}
\end{figure}
%================================================

We also consider the effect of misidentifying the cluster centre. For each cluster in the simulations, we calculate galaxy numbers, radii, etc. not only using the `true' centre, but also using the position of the second-most massive subhalo as `apparent' centre. The resulting distribution of projected offsets $R_s$ between `true' and 'apparent' centre is shown in Fig.~\ref{fig:offset_distribution} for the MS. The distribution can be approximated by a two-dimensional Gaussian distribution with
\begin{equation}
\label{eq:offset_distribution}
 \pdf(R_s)=\frac{R_s}{\sigma_s^2}\exp\left(-\frac{R_s^2}{2 \sigma_s^2}\right)
\end{equation}
and $\sigma_s=0.38h^{-1}\Mpc$, which agrees well with the offset distribution found for the maxBCG algorithm run on simulated data \citep{JohnstonEtal2007_SDSS_cluster_wl_II_arXiv}. We find a similar value, $\sigma_s=0.41h^{-1}\Mpc$, for the WMAP1 simulation, and somewhat smaller values, $\sigma_s=0.34h^{-1}\Mpc$, for the WMAP3 simulation.

When needed for our analysis, we will use a probability
\begin{equation}
\label{eq:a_priori_center_fraction}
 \tilde{p}_\text{c}(\Nint) = \frac{1+0.04 \Nint}{2.2 + 0.05 \Nint}
\end{equation}
that a cluster in the simulation with $\Nint$ ridgeline galaxies is correctly centred. Empirically, this yields roughly the same probability $p_\text{c}(\N200)$ that a cluster with measured richness $\N200$ is correctly centred as was found for the maxBCG algorithm by \citet{JohnstonEtal2007_SDSS_cluster_wl_II_arXiv}.

In our simulations, centre misidentification tends to reduce the number $\N200$ of ridgeline galaxies within $\Rgal$ for a given cluster. Consequently, the number density of clusters with $\N200$ above a given threshold is slightly decreased by centre misidentification. Another consequence of centre misidentification in our simulations is a slightly higher average cluster mass and ridgeline galaxy number $\Nint$ for a given measured richness $\N200$. All these effects may be smaller for the actual maxBCG algorithm. This algorithm disfavours identifying the cluster centre with galaxies that lead to low $\N200$ in comparison to galaxies that yield a larger cluster richness. In the following, we will thus discuss results for our simulated cluster samples in the case that centre misidentification is ignored, unless stated otherwise.

%%%%%%%%%%%%%%%%%%%%%%%%%%%%%%%%%%%%%%%%%%%%%%%%%
\subsection{The cluster samples}
\label{sec:model_cluster_samples}
%%%%%%%%%%%%%%%%%%%%%%%%%%%%%%%%%%%%%%%%%%%%%%%%%

A comparison of cluster abundances in our simulated cluster samples to observation requires knowledge of the volumes and areas of the real surveys. For the SDSS maxBCG cluster sample, we assume an effective survey area of $7400\,\degt^2$ and a redshift range of $0.1\leq z \leq 0.3$ \citep[][]{RozoEtal2009_MaxBCG_V_cosmology_arXiv}. This yields an effective survey volume of $4.3 \times10^8 h^{-3}\,\Mpc^{3}$ for the MS and WMAP1 cosmology, and $4.4 \times10^8 h^{-3}\,\Mpc^{3}$ for the WMAP3 cosmology.

For each snapshot of our simulations with redshift $0.1\leq z \leq 0.3$, we create a model cluster catalogue containing the projections of all clusters in the simulation box. We calculate the statistical properties of interest (e.g. the cluster densities or average cluster mass as a function of cluster richness) for each of these snapshot catalogues (properly taking into account the increased cluster number due to inclusion of multiple projections of each cluster). We then compute the properties of a cluster sample in a volume-limited survey with $0.1\leq z \leq 0.3$ by an average over the snapshot results, where each snapshot is weighted by its cluster abundance and its volume fraction in the survey.

The simulation cube volume $L^3=1.25 \times10^8 h^{-3}\,\Mpc^{3}$ for the MS, which is about one quarter of the SDSS maxBCG survey volume. The simulation cube is used in three different projections and at six different redshifts to construct the model cluster samples, which increases the effective sample size.\footnote{
The resulting effective sample size is difficult to quantify since the subsamples created from the different projections and redshifts are not independent.}
One can thus expect the statistical errors due to sample variance to be roughly the same for the MS model and the SDSS maxBCG cluster sample, and about eight times larger for the WMAP models (with their 64 times smaller box volume).

Where appropriate, we estimate the errors of our models due to sample variance in the following way:
We divide the simulation cube of the MS into 64 smaller cubes each having the size of the WMAP simulations. We calculate the observables for each of these subcubes separately. The standard deviation of the results from the different subcubes serves as an estimate for the statistical error of the WMAP1-A model. The statistical errors for the other models are than extrapolated from the WMAP1-A error using simple assumptions about the scaling with volume, numbers, etc.

%%%%%%%%%%%%%%%%%%%%%%%%%%%%%%%%%%%%%%%%%%%%%%%%%
\section{Results}
\label{sec:results}
%%%%%%%%%%%%%%%%%%%%%%%%%%%%%%%%%%%%%%%%%%%%%%%%%

Here, we compare the properties of clusters in our various galaxy formation models to the observed properties of clusters in the SDSS maxBCG catalogue. We investigate average cluster properties as a function of galaxy content by dividing our simulated cluster samples into bins of richness $\N200$ and luminosity $\L200$ (as was done for the maxBCG cluster sample). The average properties of the model clusters binned by $\N200$ are listed for the different galaxy formation models in Tables~\ref{tab:clusters_summary_MS}-\ref{tab:clusters_summary_WMAP3_C}. The properties of model clusters when binned by $\L200$ are shown in Tables~\ref{tab:clusters_summary_MS_L}-\ref{tab:clusters_summary_WMAP3_C_L}.

We first present results for stacked weak-lensing mass profiles and for the abundance of clusters, both as a function of galaxy richness. We then discuss the mean and scatter of cluster mass as functions of richness and luminosity. Finally, we study how well the cluster mass function and the stacked weak-lensing mass profiles can reconstructed from the cluster abundance together with the mean and scatter of the mass-richness relation.

%%%%%%%%%%%%%%%%%%%%%%%%%%%%%%%%%%%%%%%%%%%%%%%%%
\subsection{Cluster mass profiles}
\label{sec:cluster_mass_profiles}
%%%%%%%%%%%%%%%%%%%%%%%%%%%%%%%%%%%%%%%%%%%%%%%%%

For each cluster in the simulations, we compute weak-lensing mass profiles by projecting the simulation particles and the galaxies in a cuboid region of $35h^{-1}\,\Mpc$ transverse physical side length and $100h^{-1}\,\Mpc$ comoving thickness centred on the cluster. The projected particles and galaxies (assumed to contribute their stellar mass as points)\footnote{
Although particles in the simulations represent the total mass in the simulated parts of the universe, we do not compensate for the additional stellar mass, since (i) the mass in stars is very small compared to the total mass in collapsed objects, and (ii) gas physics increases the dark-matter density in the inner part of the halos compared to collisionless simulations \citep[e.g.][]{BarnesWhite1984,GnedinEtal2004}.
} are binned in annuli to compute the circularly averaged surface mass density $\Sigma(R)$ at radius $R$ from the projected cluster centre, the mean enclosed surface mass density $\bar{\Sigma}(R)$ inside $R$, and the weak-lensing mass profile
\begin{equation}
\Delta \Sigma(R) = \bar{\Sigma}(R) - \Sigma(R).
\end{equation}

The weak-lensing mass profile $\Delta \Sigma(R)$ is proportional to the average tangential shear $\EV{\gamma_\mathrm{t}}(R)$ around the projected cluster centre, and can therefore be measured using weak-lensing observations \citep[][]{SchneiderKochanekWambsganss_book}. To increase signal-to-noise, the measured tangential shear may be averaged over a sample of clusters. The resulting shear signal can then be converted to an average mass profile for the observed cluster sample.

In this section, the weak-lensing mass profiles of the simulated clusters in each snapshot are averaged in bins of $\N200$. The average profiles from each snaphot are then combined with appropriate weights to obtain a mean profile for each $\N200$-bin in a volume-limited survey. In doing this, we also take into account the effect of cluster centre misidentification, which has a considerable impact on the average profiles at radii $R<1h^{-1}\,\Mpc$ (see Sec.~\ref{sec:cluster_mass_profile_fits}).

%================================================
\begin{figure*}
\centerline{\includegraphics[width=0.8\linewidth]{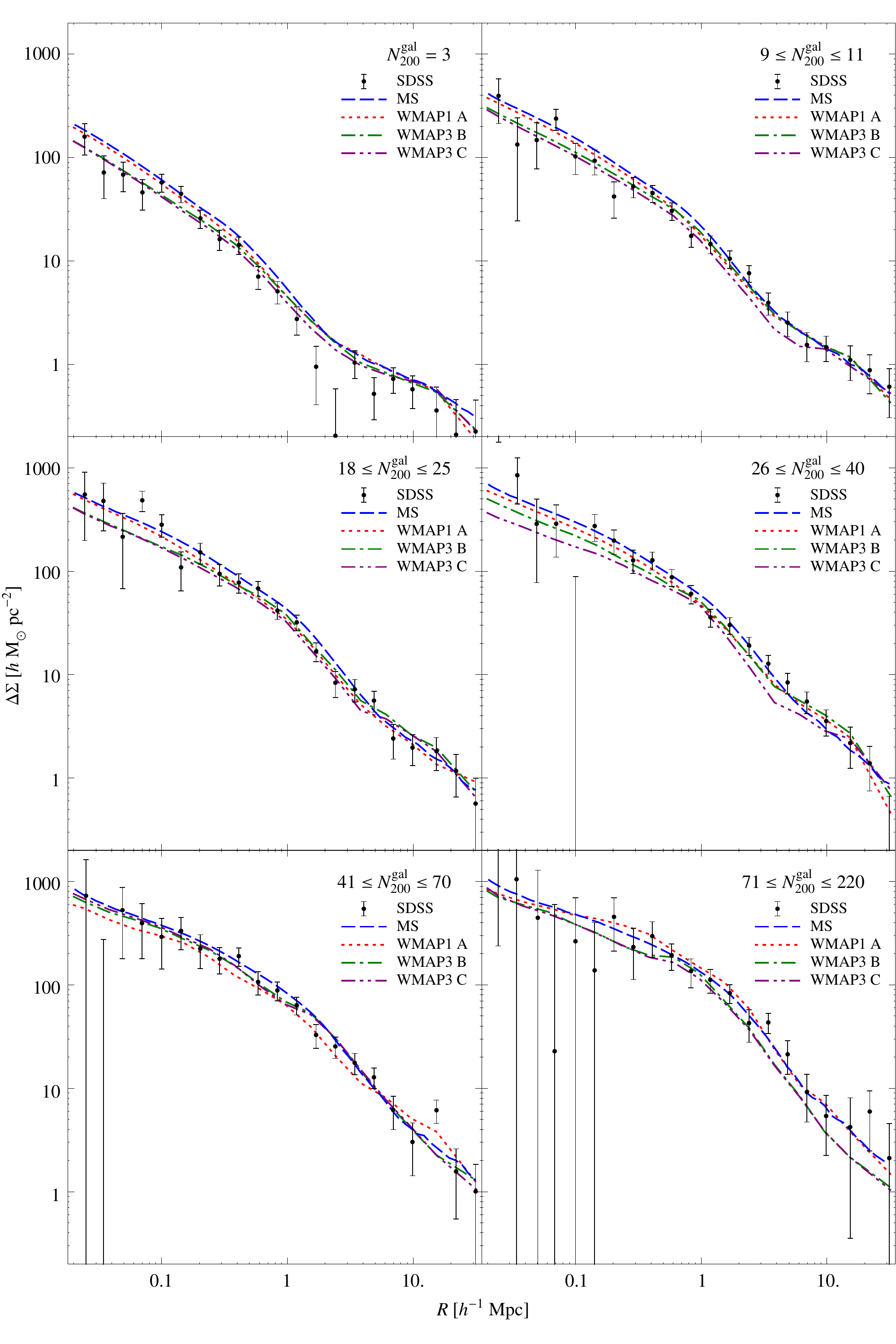}}
\caption{
\label{fig:mass_profiles}
The average weak-lensing mass profile $\Delta \Sigma(R)$ as a function of radius $R$ for clusters in different richness bins in the MS (dashed lines), WMAP1-A (dotted lines), WMAP3-B (dash-dotted lines), and the WMAP3-C model (dash-dot-dotted lines) compared to the observed profiles in the SDSS \protect\citep[][points with error bars]{SheldonEtal2009_SDSS_cluster_wl_I}.
}
\end{figure*}
%================================================

In Fig.~\ref{fig:mass_profiles}, the weak-lensing mass profiles for the simulated clusters in the MS and the WMAP models are compared to the measured profiles of maxBCG clusters in the SDSS \citep{SheldonEtal2009_SDSS_cluster_wl_I}. The simulated and observed profiles agree remarkably well in detailed shape and amplitude. This is strong evidence that our models provide a realistic description not only of the density profile and galaxy content of galaxy clusters, but also of the maxBCG cluster selection and richness measurement.

Differences between the galaxy models and the observations are small but noticeable. The simulated density profiles of the MS tend to be above the observed profiles in the poorer clusters but are an excellent fit in the rich systems. The WMAP3 model profiles fit better in the poor clusters but are mostly below the observed profiles in richer clusters. For radii $R\lesssim 1h^{-1}\Mpc$, the mass profiles of the WMAP3 models are roughly 30\% lower than in the MS/WMAP1 models. This suggests that for given richness, the maxBCG clusters are on average slightly less massive than clusters in the MS/WMAP1 simulations but slightly more massive than those in the WMAP3 simulations. We will investigate this issue in more detail in Sec.~\ref{sec:mass_richness_relation}.

%%%%%%%%%%%%%%%%%%%%%%%%%%%%%%%%%%%%%%%%%%%%%%%%%
\subsection{Cluster abundance}
\label{sec:cluster_abundance}
%%%%%%%%%%%%%%%%%%%%%%%%%%%%%%%%%%%%%%%%%%%%%%%%%

%================================================
\begin{table}
\center
  \caption{
\label{tab:cluster_densities}
The area density $n^\text{ang}_{\geq 10}$ [in $\degt^{-2}$] of clusters with richness $\N200\geq 10$ and redshift $0.1 \leq z \leq 0.3$ for the model clusters in the MS and WMAP simulations and for the SDSS maxBCG clusters \protect\citep{KoesterEtal2007_SDSS_clusters}. Considered are the cases that all cluster centres are correctly identified ($p_\text{c}=1$), and that only a fraction of clusters given by Eq.~\ref{eq:a_priori_center_fraction} is correctly centred ($p_\text{c}<1$).
}
\begin{tabular}{l l l}
\hline
\hline
        & $p_\text{c}=1$ & $p_\text{c}<1$ \\
\hline
SDSS    &                & 1.8            \\
MS      & 1.7            & 1.4            \\
WMAP1-A & 1.8            & 1.6            \\
WMAP3-B & 0.6            & 0.5            \\
WMAP3-C & 0.9            & 0.7            \\
\hline
\end{tabular}
\end{table}
%================================================

Table~\ref{tab:cluster_densities} compares the area density $n^\text{ang}_{\geq 10}$ of clusters with richness $\N200\geq10$ and redshift $0.1 \leq z \leq 0.3$ in our simulated surveys to the observed abundance of maxBCG clusters in the SDSS. If we ignore misidentification of the cluster centres, the MS and the WMAP1-A model yield cluster abundances very similar to the real maxBCG cluster sample. In contrast, the cluster abundances for the WMAP3 models are a factor 2-3 lower than that observed in the SDSS. The lowest abundance is found for the WMAP3-B model.

To obtain an estimate for the statistical error on the cluster density, we employ the subsampling method described in Sec.~\ref{sec:model_cluster_samples}: We use the MS model to create subsamples with sizes equal to the WMAP1-A model. From these subsamples, we estimate a standard deviation for the area density of clusters of 25\% for the WMAP1-A model. Taking into account the lower cluster densities in the WMAP3 models, a slightly larger statistical error of 30-40\% can be assumed for these. A simple extrapolation of the WMAP1-A error yields a much smaller error of 3\% for the cluster density in the MS model.

Taking into account centre misidentification in our simulations reduces the cluster abundances by $\approx 20\%$ and thus increases the discrepancy between the WMAP3 models and the SDSS cluster sample. The abundance is then also somewhat low even in the MS case, suggesting, as noted above, that our procedure for modelling the effects of centre misidentification may overestimate the effect in the real maxBCG catalogues.

%================================================
\begin{figure}
\centerline{\includegraphics[width=1\linewidth]{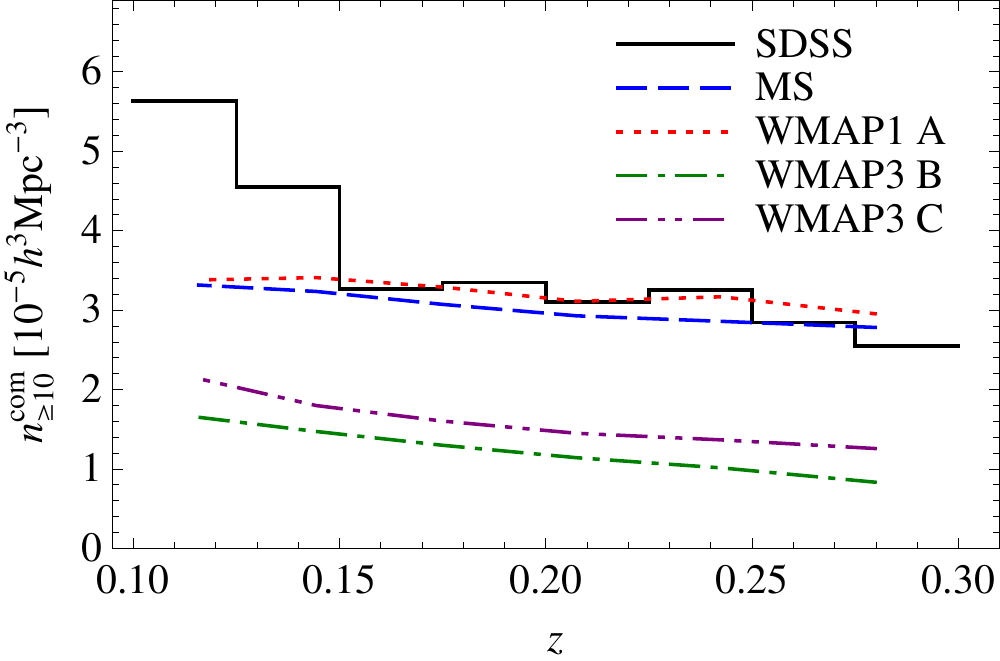}}
\caption{
\label{fig:cluster_density_vs_redshift}
The comoving abundance $n_{\geq10}^\text{com}(z)$ of clusters with richness $\N200\geq 10$ as a function of redshift $z$ in the SDSS maxBCG catalogue \protect\citep{KoesterEtal2007_SDSS_clusters} and in our simulated cluster catalogues based on the MS and the WMAP simulations.
}
\end{figure}
%================================================

The redshift dependence of the comoving cluster abundance is shown in Fig.~\ref{fig:cluster_density_vs_redshift}. All our models show a slight decrease of the comoving cluster density with increasing redshift. The statistical errors on the comoving cluster abundance estimated with the subsampling method are similar to the errors on the cluster area density, i.e. 25-40\% for the WMAP models and $\sim3\%$ for the MS. For $z>0.15$, the abundance in the MS and WMAP1-A model agree very well with the maxBCG results. In contrast, there is a much larger observed density at low redshifts $z<0.15$, which is not seen in the models and presumably reflects nearby large-scale structure such as the SDSS Great Wall.

%================================================
\begin{figure}
\centerline{\includegraphics[width=1\linewidth]{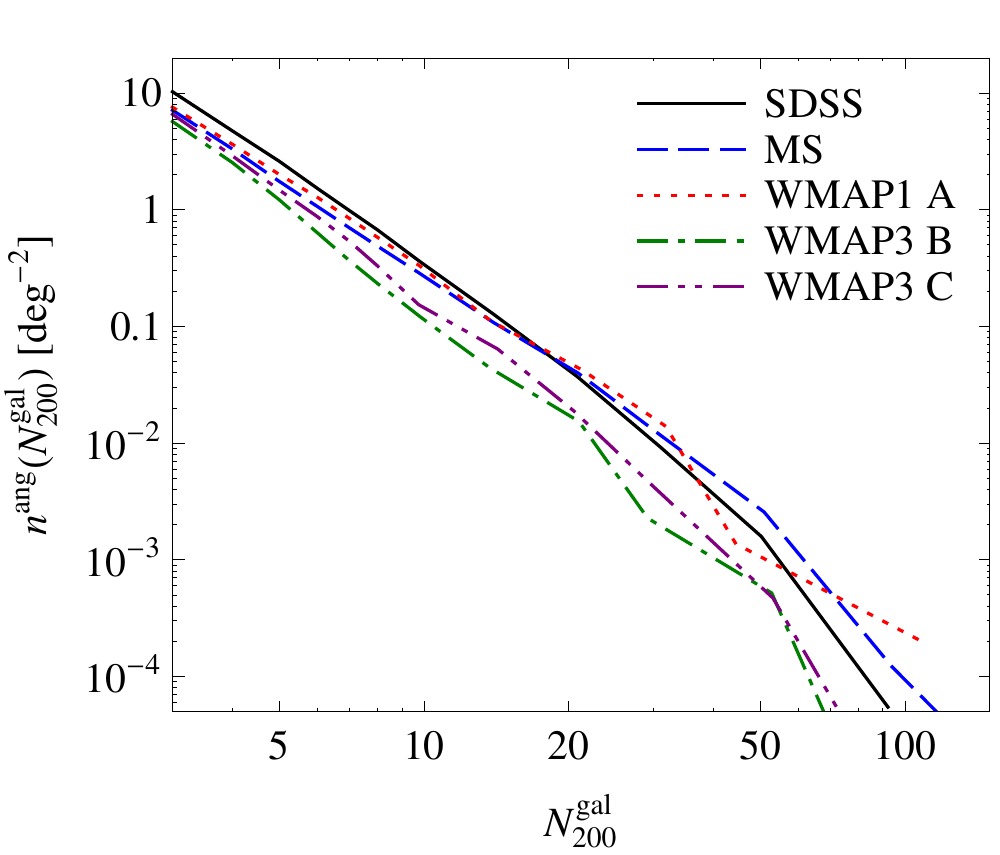}}
\caption{
\label{fig:cluster_histogram}
The surface density $n^\text{ang}(\N200)$ of clusters with redshift $0.1 \leq z \leq 0.3$ as a function of richness $\N200$. Counts in the SDSS maxBCG cluster sample \protect\citep{KoesterEtal2007_SDSS_clusters} are compared to counts in our simulated cluster catalogues for the MS and the WMAP simulations.
}
\end{figure}
%================================================

The dependence of the cluster counts on richness $\N200$ is illustrated in Fig.~\ref{fig:cluster_histogram}. Cluster abundances in the MS and WMAP1-A models are lower than in the SDSS for $\N200 < 20$, but exceed the observed abundances for $\N200\geq20$. Abundances in the WMAP3 models are always lower than in the MS and WMAP1-A model and below the observations. For $\N200\geq9$, the cluster abundances in the WMAP3 models are 2-20 times lower than the abundances in the WMAP1-A model and 2-5 times lower than the observed abundances. Hence, the differences are always larger than the statistical errors inferred from the subsampling (which are $\sim$10-100\% for the WMAP models and $\sim$1-10\% for the MS and SDSS). The differences are largest in the high-$\N200$ tail of the distribution.

The low cluster abundance for the WMAP3 models in comparison to the MS and WMAP1-A models, in particular for large $\N200$, are a reflection of the different cosmologies. Rich clusters have massive dark matter halos, and there are far fewer massive halos in the WMAP3 cosmology than in the WMAP1 cosmology. This is mainly due to the lower value of $\sigma_8$. The higher star formation efficiency in the WMAP3 models does not sufficiently enhance the number of bright ridgeline galaxies per unit mass to compensate for the decrease in the number of massive halos. As a result, there are fewer rich clusters in the WMAP3 models than in the MS and WMAP1-A models.

The observed abundance of rich clusters is slightly lower than the abundance predicted for the MS and much larger than the abundance predicted by the WMAP3 models. This suggests that $0.72<\sigma_8<0.9$ for our Universe, with $\sigma_8$ probably closer to 0.9 than to 0.72. This is consistent with some recent estimates [e.g. $\sigma_8 = 0.80 \pm 0.02$ by \citealp[][]{LesgourguesEtal2007},
$\sigma_8=0.81 \pm 0.03$ by \citealp[][]{KomatsuEtal2009}, and $\sigma_8 = (0.83\pm 0.03)(\Omega_\mathrm{M}/0.25)^{-0.41}$ by \citealp[][]{RozoEtal2009_MaxBCG_V_cosmology_arXiv}].

The results for the abundances of rich clusters alone are not sufficient to definitely conclude that $0.72<\sigma_8<0.9$. For example, problems with the modelling of the ridgeline galaxies and their identification could have lead to inaccurate estimates for the cluster richness and thus to incorrect cluster abundances. However, there is complementary evidence from the mass-richness relation, which we discuss in Sec.~\ref{sec:mass_richness_relation} and \ref{sec:mass_function}.

%%%%%%%%%%%%%%%%%%%%%%%%%%%%%%%%%%%%%%%%%%%%%%%%%
\subsection{Mass-richness relation}
\label{sec:mass_richness_relation}
%%%%%%%%%%%%%%%%%%%%%%%%%%%%%%%%%%%%%%%%%%%%%%%%%

%================================================
\begin{figure}
\centerline{\includegraphics[width=1\linewidth]{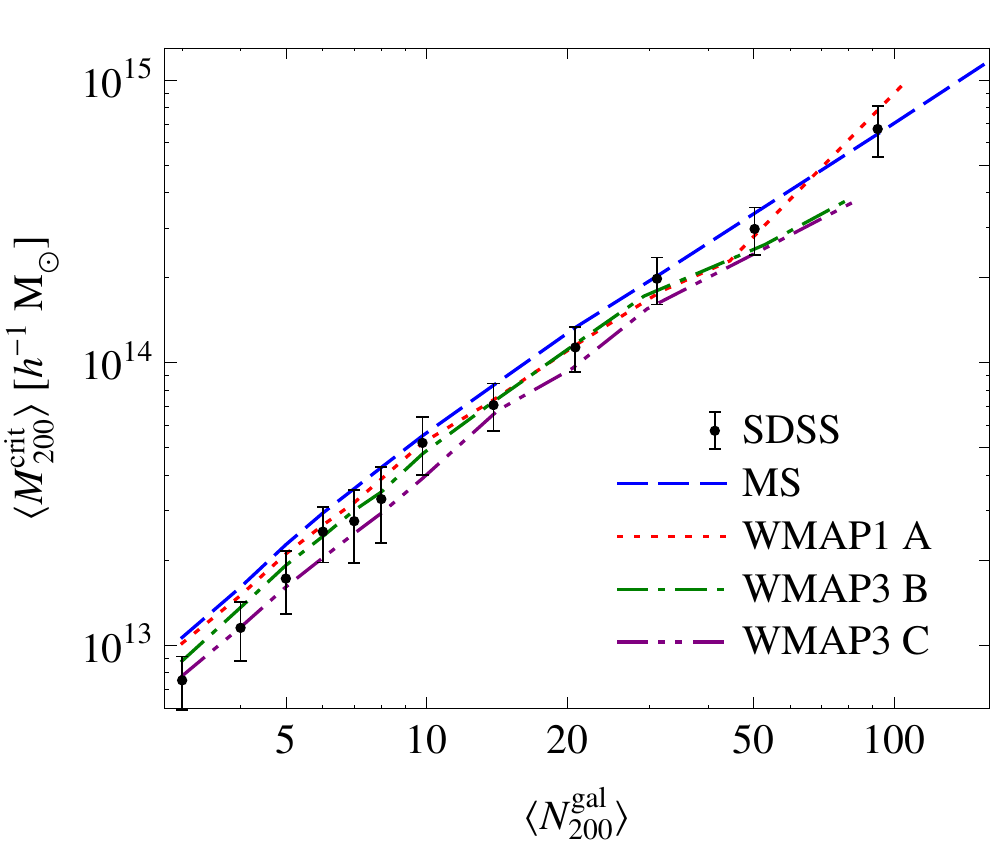}}
\caption{
\label{fig:gal_number_vs_crit_cluster_mass}
The average cluster mass $\ev{\Mcrit}$ vs. richness $\ev{\N200}$ relations for cluster catalogues from the MS and WMAP simulations are compared to the relation derived from the SDSS maxBCG catalogues by \protect\citet{JohnstonEtal2007_SDSS_cluster_wl_II_arXiv}.
}
\end{figure}
%================================================

The weak-lensing mass profiles discussed in Sec.~\ref{sec:cluster_mass_profiles} can be used to estimate spherical-overdensity cluster masses. This can be done, e.g., by a non-parametric conversion of the weak-lensing mass profiles into average 3D density profiles \citep[][]{JohnstonEtal2007}, or by fitting a parametrised model of the average cluster density to the shear data \citep[][]{JohnstonEtal2007_SDSS_cluster_wl_II_arXiv}.

In Fig.~\ref{fig:gal_number_vs_crit_cluster_mass}, average cluster masses $\ev{\Mcrit}$ in our simulated catalogues\footnote{
We use the masses measured directly from the matter distribution in the simulations. As discussed in Sec.~\ref{sec:cluster_mass_profile_fits}, these are consistent with the masses obtained from parametric fits to the weak-lensing mass profiles.
}
are shown as a function of cluster richness $\ev{\N200}$ and are compared to the corresponding relation for SDSS maxBCG clusters by \citet{JohnstonEtal2007_SDSS_cluster_wl_II_arXiv}, who calculated the cluster masses by fitting parametric models to the observed lensing signal.\footnote{We multiplied the masses given in \citet{JohnstonEtal2007_SDSS_cluster_wl_II_arXiv} by a factor $1.18\pm0.06$, which is a photo-$z$ bias correction advocated by \citet{MandelbaumEtal2008} and \citet{RozoEtal2009_MaxBCG_III_scatter}.} Remarkably, all our simulations reproduce the observed mass-richness relation within $\sim30\%$ over the two orders of magnitude spanned by the SDSS clusters. This is corroborates our finding in Sec.~\ref{sec:cluster_mass_profiles} that our models provide an adequate description of the statistical properties of optically selected galaxy clusters.

At given richness, the differences in cluster density profiles between models and observations (see Fig.~\ref{fig:mass_profiles}) imply differences in mean cluster mass. For $\N200<10$, the MS yields cluster masses that are 30-40\% higher than the SDSS maxBCG cluster masses derived by \citet{JohnstonEtal2007_SDSS_cluster_wl_II_arXiv}. For $\N200\geq10$, the MS masses are up to 20\% higher than the SDSS maxBCG masses. The cluster masses of the WMAP1-A model are comparable to those of the MS, but seem to be affected by sampling noise for large $\ev{\N200}$. The cluster masses in the WMAP3-B model are lower than those for the MS model by 20-30\%, and fall below the SDSS cluster masses at large richness. Among our models, the lowest cluster masses are found for the WMAP3-C model, where the values are always smaller than those in the SDSS.

The mass-richness relation for the WMAP1-A model differs by up to 30\% from the relation for the MS model (which is much less affected by sampling noise due to its 64 times larger volume). Using subsamples created from the MS as described in Sec.~\ref{sec:model_cluster_samples}, we estimate a standard deviation for the binned cluster mass of 2\% for $\N200=3$ and 20\% for $71\leq\N200\leq220$ for the WMAP1-A model. A similar statistical error can be assumed for the WMAP3 models. This is consistent with the differences between the MS and WMAP1-A models being solely due to sampling noise.

%================================================
\begin{figure}
\centerline{\includegraphics[width=1\linewidth]{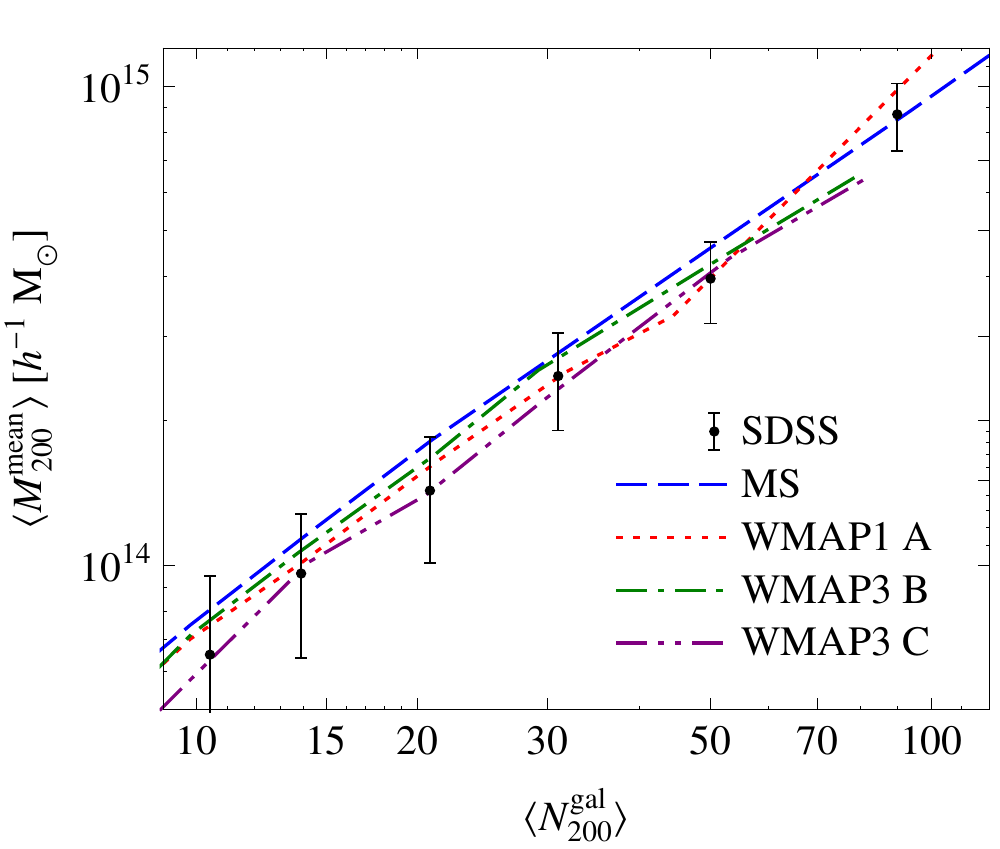}}
\caption{
\label{fig:gal_number_vs_mean_cluster_mass}
The average cluster mass $\ev{\Mmean}$ vs. richness $\ev{\N200}$ relations for  cluster catalogues from our MS and WMAP simulations are compared to the SDSS relation given by \protect\citet{ReyesEtal2008}.
}
\end{figure}
%================================================

Another analysis of the weak-lensing data for the SDSS maxBCG cluster sample has been performed by \citet{ReyesEtal2008}. In Fig.~\ref{fig:gal_number_vs_mean_cluster_mass}, we compare the average cluster masses $\ev{\Mmean}$ of our simulated clusters to their results as a function of richness. As for $\ev{\Mcrit}$, the average cluster masses $\ev{\Mmean}$ in the MS model are up to 30\% larger than the SDSS maxBCG cluster masses, whereas the cluster masses of the WMAP3 models are comparable to those in the SDSS.

%================================================
\begin{table}
\center
\caption{
\label{tab:N_200_vs_M_200crit_fit_params}
The best-fit parameters (calculated from a least-squares fit of $\log_{10}\ev{\N200}$ against $\log_{10}\ev{\Mcrit}$) for the mass-richness relation~\eqref{eq:N_200_vs_M_200_fit} of simulated clusters with $\N200\geq9$ is compared to the best-fitting parameters for the SDSS maxBCG clusters with masses measured by \protect\citet{JohnstonEtal2007_SDSS_cluster_wl_II_arXiv}.
}
\begin{tabular}{l c c c c}
\hline
\hline
        & $M^\text{crit}_{200|20}$ [$h^{-1} \Msolar$] & $\alpha_N^\text{crit}$ \\
\hline
SDSS    & $(1.12 \pm 0.03)\times10^{14}$              & $1.14 \pm 0.04$        \\
MS      & $(1.24 \pm 0.01)\times10^{14}$              & $1.09 \pm 0.01$        \\
WMAP1-A & $(1.11 \pm 0.08)\times10^{14}$              & $1.21 \pm 0.09$        \\
WMAP3-B & $(1.05 \pm 0.04)\times10^{14}$              & $0.97 \pm 0.05$        \\
WMAP3-C & $(0.92 \pm 0.04)\times10^{14}$              & $1.05 \pm 0.05$        \\
\hline
\end{tabular}
\end{table}
%================================================

%================================================
\begin{table}
\center
\caption{
\label{tab:N_200_vs_M_200mean_fit_params}
The best-fit parameters (calculated from a least-squares fit of $\log_{10}\ev{\N200}$ against $\log_{10}\ev{\Mmean}$) for the mass-richness relation~\eqref{eq:N_200_vs_M_200_fit} of simulated clusters with $\N200\geq9$ is compared to the best-fit parameters for the SDSS maxBCG clusters with masses measured by \protect\citet{ReyesEtal2008}.
}
\begin{tabular}{l c c c c}
\hline
\hline
        & $M^\text{mean}_{200|20}$ [$h^{-1} \Msolar$] & $\alpha_N^\text{mean}$ \\
\hline
SDSS    & $(1.42 \pm 0.03)\times10^{14}$              & $1.19 \pm 0.03$        \\
MS      & $(1.68 \pm 0.02)\times10^{14}$              & $1.08 \pm 0.01$        \\
WMAP1-A & $(1.53 \pm 0.08)\times10^{14}$              & $1.19 \pm 0.06$        \\
WMAP3-B & $(1.58 \pm 0.04)\times10^{14}$              & $1.06 \pm 0.03$        \\
WMAP3-C & $(1.38 \pm 0.04)\times10^{14}$              & $1.13 \pm 0.04$        \\
\hline
\end{tabular}
\end{table}
%================================================

The mass-richness relations shown in Fig.~\ref{fig:gal_number_vs_crit_cluster_mass} and \ref{fig:gal_number_vs_mean_cluster_mass} suggest a power law, although with a steeper slope for $\ev{\N200}<10$ than for $\ev{\N200}\gtrsim10$. Here, we fit the mean mass-richness relation for clusters with $\N200 \geq 9$ by:
\begin{equation}
\label{eq:N_200_vs_M_200_fit}
 \Mcritmean(\N200) = M^\text{crit/mean}_{200|20}
 \left(\frac{\N200}{20}\right)^{\alpha_N^\text{crit/mean}}.
\end{equation}
The best-fit parameters for our simulated catalogues are compared to those for the SDSS in Tables~\ref{tab:N_200_vs_M_200crit_fit_params} and \ref{tab:N_200_vs_M_200mean_fit_params}.

The lower cluster masses in the WMAP3 models than in the MS and WMAP1-A models again reflect the different cosmologies. The less evolved dark matter structure in the WMAP3 cosmology requires more efficient star formation in order to match observed galaxy numbers. This results in more ridgeline galaxies in a dark matter halo of given mass. Consequently, for a given richness, halos are less massive in the WMAP3 models than in the MS and WMAP1-A models.

Except for the largest richness bin, observed cluster masses are smaller than in the MS model. This suggests a normalisation $\sigma_8 < 0.9$ for our Universe \citep[again consistent with the recent estimate $\sigma_8=0.81 \pm 0.03$ by][]{KomatsuEtal2009}.

%%%%%%%%%%%%%%%%%%%%%%%%%%%%%%%%%%%%%%%%%%%%%%%%%
\subsection{Mass-luminosity relation}
\label{sec:mass_luminosity_relation}
%%%%%%%%%%%%%%%%%%%%%%%%%%%%%%%%%%%%%%%%%%%%%%%%%

%================================================
\begin{figure}
\centerline{\includegraphics[width=1\linewidth]{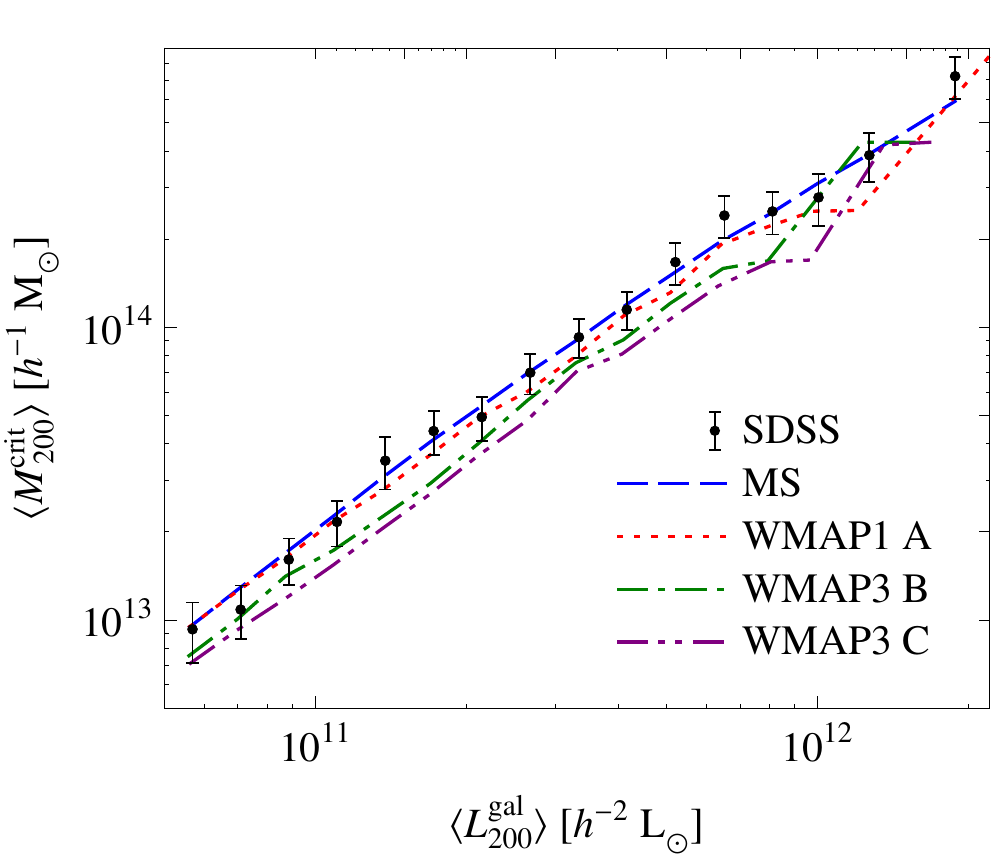}}
\caption{
\label{fig:gal_lum_vs_crit_cluster_mass}
Average cluster mass $\ev{\Mcrit}$ vs. the total $i$-band luminosity $\ev{\L200}$ of ridgeline galaxies within $\Rgal$. Results for our MS and WMAP simulations are compared to SDSS results based on cluster masses by \protect\citet{JohnstonEtal2007_SDSS_cluster_wl_II_arXiv}.
}
\end{figure}
%================================================

%================================================
\begin{table}
\center
\caption{
\label{tab:L_200_vs_M_crit_200_fit_params}
The best-fit parameters for the mass-luminosity relation~\eqref{eq:L_200_vs_M_200_fit} in our simulations (calculated from a
least-squares fit of $\log_{10}\ev{\L200}$ vs. $\log_{10}\ev{\Mcrit}$) are compared to the values by \protect\citet{JohnstonEtal2007_SDSS_cluster_wl_II_arXiv} for the SDSS maxBCG cluster sample.
}
\begin{tabular}{l c c}
\hline
\hline
  & $M^\text{crit}_{200|40}$ [$h^{-1} \Msolar$] & $\alpha_L^\text{crit}$ \\
\hline
SDSS            & $(1.09 \pm 0.03)\times 10^{14}$  & $1.23 \pm 0.03$ \\
MS              & $(1.13 \pm 0.02)\times 10^{14}$  & $1.18 \pm 0.01$ \\
WMAP1-A         & $(1.00 \pm 0.04)\times 10^{14}$  & $1.16 \pm 0.04$ \\
WMAP3-B         & $(0.89 \pm 0.02)\times 10^{14}$  & $1.16 \pm 0.02$ \\
WMAP3-C         & $(0.81 \pm 0.02)\times 10^{14}$  & $1.17 \pm 0.02$ \\
\hline
\end{tabular}
\end{table}
%================================================

The mass-luminosity relation computed by binning our simulated cluster samples in luminosity $\L200$ is shown in Fig.~\ref{fig:gal_lum_vs_crit_cluster_mass}. For given luminosity, the MS model yields cluster masses very similar to the SDSS masses of \citet{JohnstonEtal2007_SDSS_cluster_wl_II_arXiv}, whereas the WMAP3 model produces mean cluster masses that are generally smaller than the SDSS masses. The best-fit parameters for the mass-luminosity relation
\begin{equation}
\label{eq:L_200_vs_M_200_fit}
 \Mcrit(\L200) = M^\text{crit}_{200|40} \left(\frac{\L200}{4\times10^{11}h^{-2}\Lsolar}\right)^{\alpha_L^\text{crit}}
\end{equation}
are listed in Table~\ref{tab:L_200_vs_M_crit_200_fit_params}.

The differences in the mass-luminosity relation between the various galaxy formation models can be explained in the same way as the differences in the mass-richness relation. The higher star formation efficiency in the WMAP3 models creates more bright ridgeline galaxies in a halo of a given mass $\Mcrit$ than in the MS and WMAP1-A models. This leads to lower average cluster masses at given luminosity $\L200$.

%%%%%%%%%%%%%%%%%%%%%%%%%%%%%%%%%%%%%%%%%%%%%%%%%
\subsection{The scatter in the mass-richness relation}
\label{sec:mass_richness_scatter}
%%%%%%%%%%%%%%%%%%%%%%%%%%%%%%%%%%%%%%%%%%%%%%%%%

%================================================
\begin{figure}
\centerline{\includegraphics[width=1\linewidth]{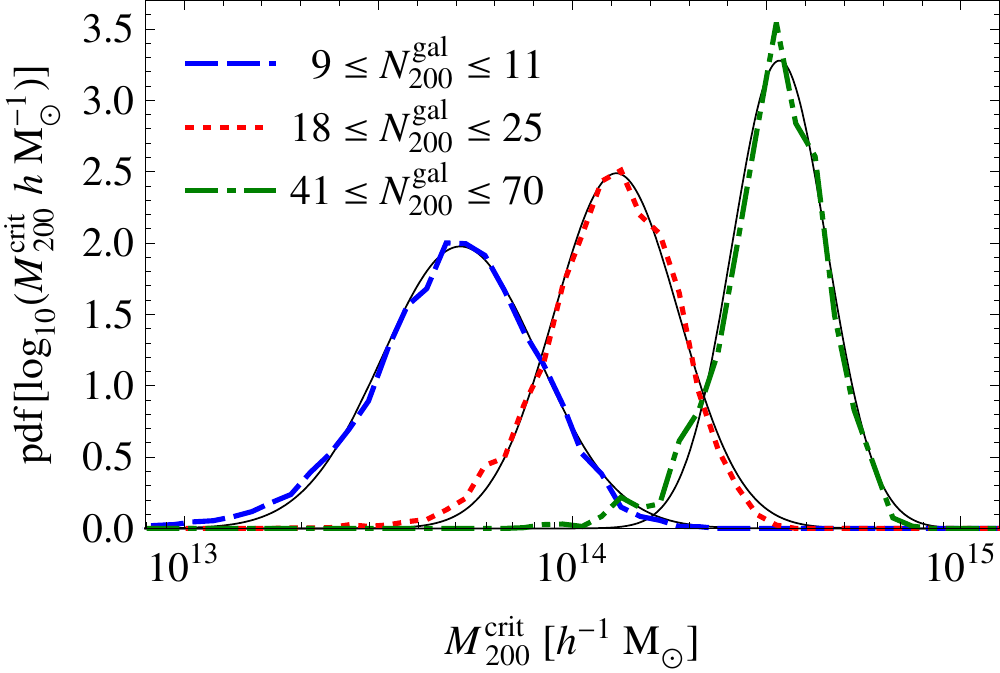}}
\caption{
\label{fig:log_mass_pdf}
The distribution of the logarithm of cluster mass $\log_{10}(\Mcrit)$ in various bins of richness $\N200$. Shown are distributions for the MS model (dashed/dotted lines) and fits of these distributions to a normal distribution (solid lines).
}
\end{figure}
%================================================

To compute a cluster mass function from cluster counts and mean masses as a function of richness, one needs to model the scatter in mass at each richness. In Fig.~\ref{fig:log_mass_pdf}, distributions of the logarithm of cluster mass $\log_{10}(\Mcrit)$ are shown for the MS model for several bins in $\N200$. These distributions are well described by gaussians.

For our various galaxy formation models, the standard deviation $\sigma_{\log_{10}(\Mcrit)}$ of the scatter in the logarithm of cluster mass $\log_{10}(\Mcrit)$ at given $\N200$ is listed for various $\N200$- and $\L200$-bins is in Tables~\ref{tab:clusters_summary_MS}-\ref{tab:clusters_summary_WMAP3_C_L}. The scatter decreases with increasing $\N200$ or $\L200$ and tends to be larger at given $\N200$ than at the corresponding $\L200$.

Our values for the scatter in the mass-richness relation are in good agreement with those found empirically by \citet{RozoEtal2009_MaxBCG_III_scatter} for the maxBCG cluster sample (using X-ray luminosities as an additional mass proxy): We find $\sigma_{\log_{10}(\Mcrit)}\approx0.15\,$-$\,0.2$ for the richness bins with  $\N200\geq9$, while \citet{RozoEtal2009_MaxBCG_III_scatter} find $\sigma_{\log_{10}(\Mcrit)}\approx0.20\pm0.09$ for $\N200\approx40$. Moreover, our values are consistent with the scatter in the velocity dispersion-richness relation derived by \citet{BeckerEtal2007} for the maxBCG clusters, if centre misidentification is taken into account. Note that there may be additional effects that increase the observed scatter, but are not modelled here.

%%%%%%%%%%%%%%%%%%%%%%%%%%%%%%%%%%%%%%%%%%%%%%%%%
\subsection{The mass function}
\label{sec:mass_function}
%%%%%%%%%%%%%%%%%%%%%%%%%%%%%%%%%%%%%%%%%%%%%%%%%

%================================================
\begin{figure}
\centerline{\includegraphics[width=1\linewidth]{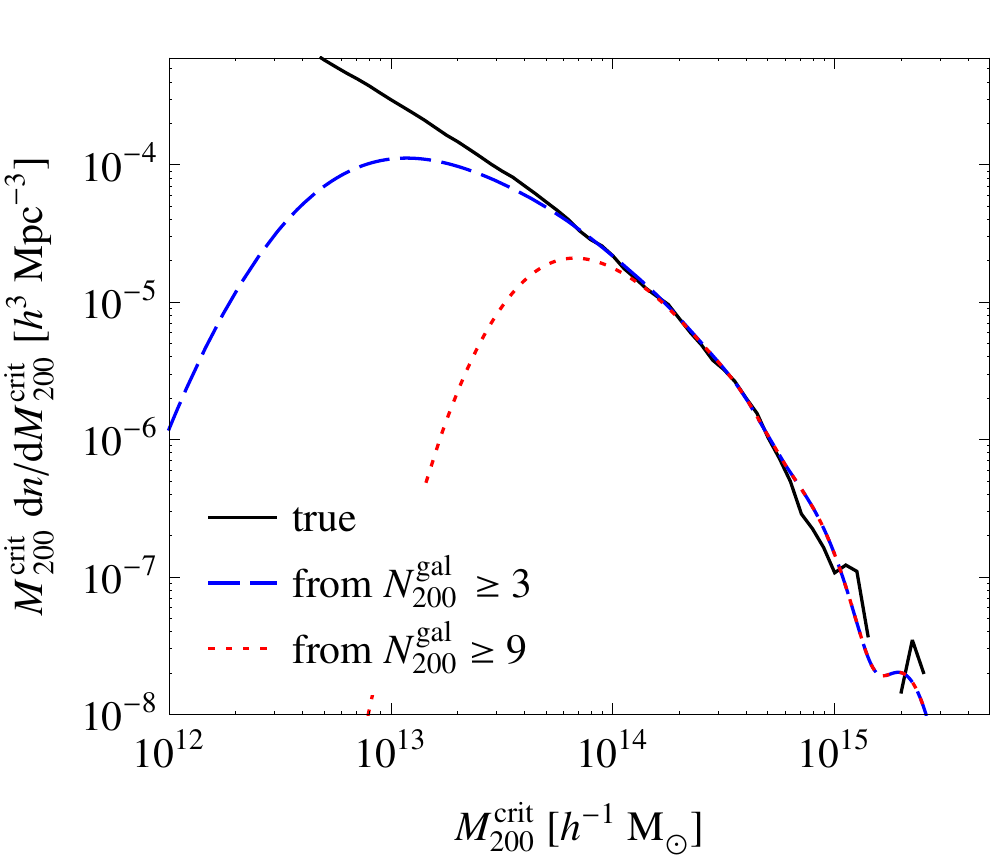}}
\caption{
\label{fig:mass_function}
The differential abundance $\diff{n}(\Mcrit)/\diff{\Mcrit}$ as a function of mass $\Mcrit$ for clusters with redshift $0.1 \leq z \leq 0.3$ in the MS: directly measured from the simulation (solid line) and reconstructed from the richness bins with $\N200\geq 3$ (dashed line) or $\N200\geq 9$ (dotted line).
}
\end{figure}
%================================================

When the cluster abundance and the mass distribution at each richness is known, it is straightforward to reconstruct the cluster mass function \citep[][]{RozoEtal2009_MaxBCG_III_scatter}. The differential cluster number density (or differential mass function) is then given by a sum over all richness bins:
\begin{equation}
  \diff{n}(\Mcrit)/\diff{\Mcrit} = \sum_{i=1}^{N_\text{bins}} n_i
  \pdf_i(\Mcrit),
\end{equation}
where $n_i$ denotes the space density and $\pdf_i(\Mcrit)$ the mass distribution for clusters in richness bin $i$.

The results of Sec.~\ref{sec:mass_richness_scatter} show that the mass distributions $\pdf_i(\Mcrit)$ at each richness can be approximated by a log-normal distribution with mean $\EV{\Mcrit}$ and scatter $\sigma_{log_{10}(\Mcrit)}$ given by the values in Table~\ref{tab:clusters_summary_MS}-\ref{tab:clusters_summary_WMAP3_C}. As Fig.~\ref{fig:mass_function} illustrates for the MS model, the reconstructed mass function matches the true mass function well for $\Mcrit \gtrsim 2\times10^{14}$. For smaller masses, the richness-selected cluster sample becomes incomplete in mass, and thus the reconstruction fails.

%================================================
\begin{figure}
\centerline{\includegraphics[width=1\linewidth]{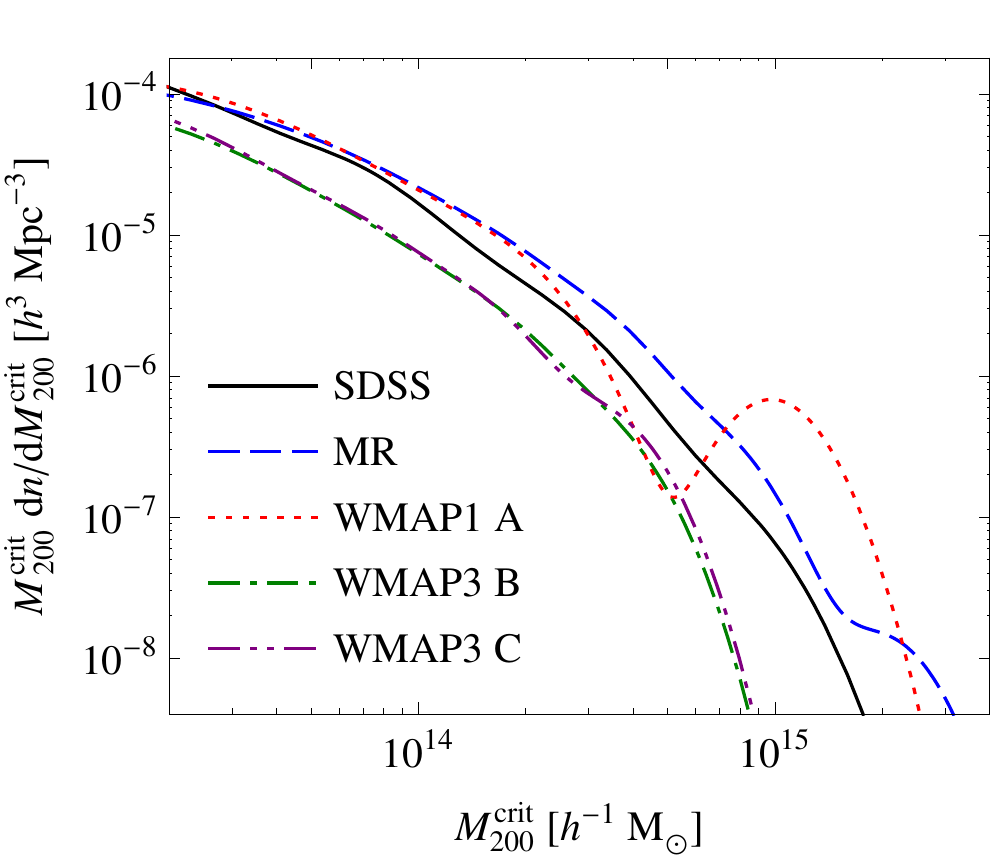}}
\caption{
\label{fig:reconstructed_mass_function}
The differential abundance $\diff{n}(\Mcrit)/\diff{\Mcrit}$ as a function of mass $\Mcrit$ for clusters with redshift $0.1 \leq z \leq 0.3$ reconstructed from the masses and abundances of clusters with richness $\N200\geq3$.  Compared are the results for MS and WMAP simulations and the SDSS (calculated from the abundances by \citealp{KoesterEtal2007_SDSS_clusters} and \protect\citealp{SheldonEtal2009_SDSS_cluster_wl_I}, the masses by \protect\citealp{JohnstonEtal2007_SDSS_cluster_wl_II_arXiv}, and the scatter by \protect\citealp{RozoEtal2009_MaxBCG_III_scatter}).
}
\end{figure}
%================================================

The reconstructed cluster mass functions for the different galaxy models and the SDSS are compared in Fig.~\ref{fig:reconstructed_mass_function}. 
The figure illustrates clearly why we expect a $\Lambda$CDM cosmology with normalisation $0.72<\sigma_8<0.9$ to provide a better fit to the SDSS cluster data than the models considered here.
The values for the WMAP3 models are always much smaller those reconstructed from the SDSS, while the MS yields values above the observations. The reconstructed cluster mass function for the WMAP1 model generally follows the MS results, but is visibly affected by sampling noise for larger cluster masses.

Since the cluster mass function can be recovered from the cluster abundances and cluster masses as functions of richness, these quantities cannot vary independently if the cluster mass function is fixed. Different assumptions about the galaxy formation physics or the richness measurements that, for given richness, lead to higher cluster abundances will also yield lower cluster masses (and vice versa).
Thus, the abundance-richness relation discussed in Sec.~\ref{sec:cluster_abundance} and the mass-richness relation discussed in Sec.~\ref{sec:mass_richness_relation} provide complementary information on the cosmology.

%%%%%%%%%%%%%%%%%%%%%%%%%%%%%%%%%%%%%%%%%%%%%%%%%
\subsection{Fits to the cluster mass profiles}
\label{sec:cluster_mass_profile_fits}
%%%%%%%%%%%%%%%%%%%%%%%%%%%%%%%%%%%%%%%%%%%%%%%%%

The results of Sec.~\ref{sec:mass_richness_scatter} justify the use of a log-normal mass distribution for fits to observed mass profiles \citep[e.g., by][]{JohnstonEtal2007_SDSS_cluster_wl_II_arXiv,ReyesEtal2008}. Here, we illustrate that one can indeed obtain a good fit to the simulated mean mass profiles of clusters with assumptions similar to those used, e.g., by \citet{JohnstonEtal2007_SDSS_cluster_wl_II_arXiv}.

%================================================
\begin{figure}
\centerline{\includegraphics[width=0.8\linewidth]{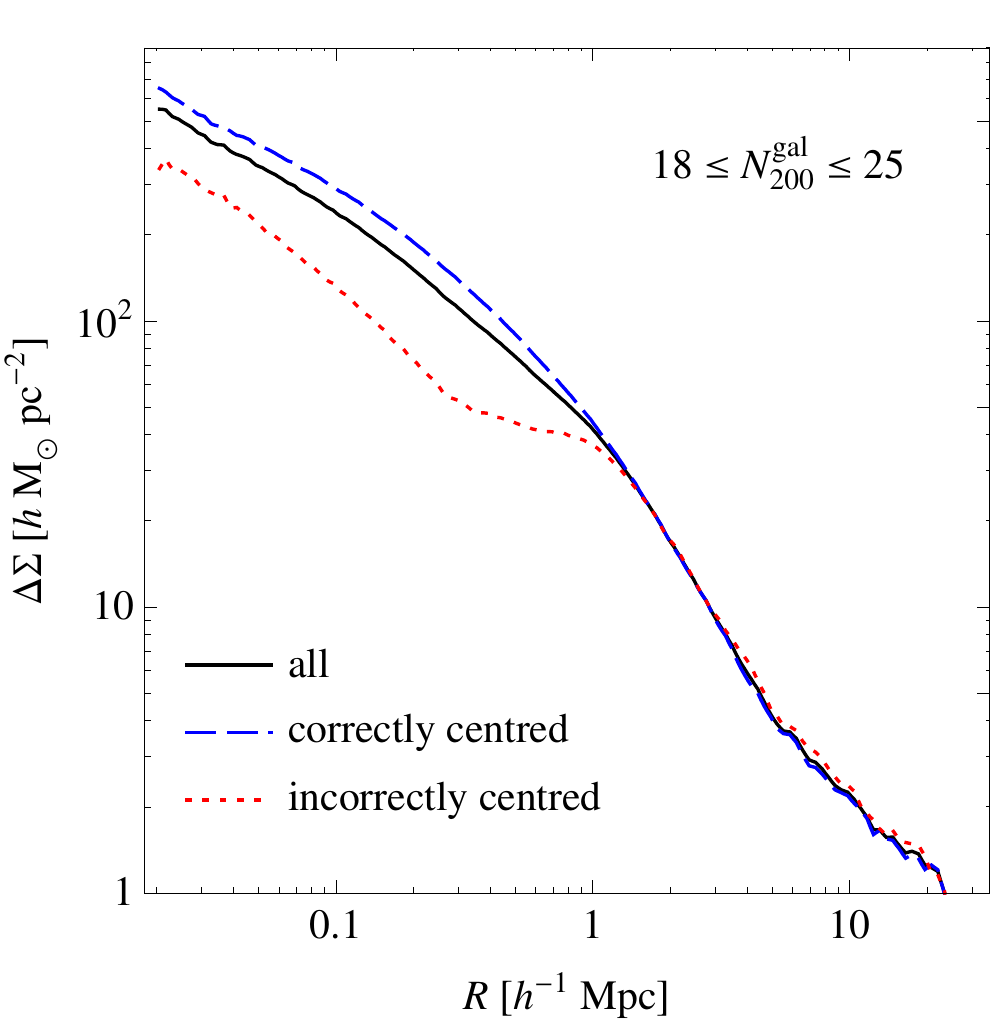}}
\caption{
\label{fig:center_vs_off_center_mass_profiles}
Comparison of the average weak-lensing mass profiles $\Delta \Sigma(R)$ as a function of radius $R$ for all clusters (solid line), for the correctly centred clusters (dashed line), and for the incorrectly centred clusters (dashed line) in the MS.
}
\end{figure}
%================================================

The mass profiles discussed in Sec.~\ref{sec:cluster_mass_profiles} are a mixture of correctly and incorrectly centred clusters. In Fig.~\ref{fig:center_vs_off_center_mass_profiles}, the simulated average mass profiles of all clusters are compared to the profiles of correctly and incorrectly centred clusters. The profiles agree well for large radii, but differ significantly below a certain radius $R \approx 0.5h^{-1}\,\Mpc$ for clusters with richness $\N200=3$, and $R\approx 1h^{-1}\,\Mpc$ for clusters with $\N200\geq9$.

As Fig.~\ref{fig:center_vs_off_center_mass_profiles} illustrates, centre misidentification has a significant impact on the average cluster mass profiles and thus needs to be taken into account in profile fits. Since stacking weak-lensing mass profiles is linear, we can discuss the contributions from correctly and incorrectly centred clusters separately. A weighted average of fits to these two components constitutes a fit to the average mass profile of all clusters in a richness bin.

We assume that the average mass profile of the correctly centred clusters in a richness bin consists of a central galaxy component, which we model as point mass, a mean dark matter halo modelled as an average over spherical NFW profiles \citep{NavarroFrenkWhite1997}, and a contribution from neighbouring mass concentrations. A log-normal distribution with mean $\EV{\Mcrit}$ and standard deviation $\sigma_{log_{10}(\Mcrit)}$ given by Table~\ref{tab:clusters_summary_MS} is assumed for the halo masses. Furthermore, we assume that the concentration $c$ of halos with mass $\Mcrit$ follows a log-normal distribution with mean
\begin{equation}
  \EV{c}(\Mcrit) = 4.67\left(\frac{\Mcrit}{10^{14}h^{-1}\Msolar}\right)^{-0.11}
\end{equation}
and standard deviation $\sigma_{\log_{10}(c)}=0.15$ \citep[][]{NetoEtal2007}.

%================================================
\begin{figure}
\centerline{\includegraphics[width=0.8\linewidth]{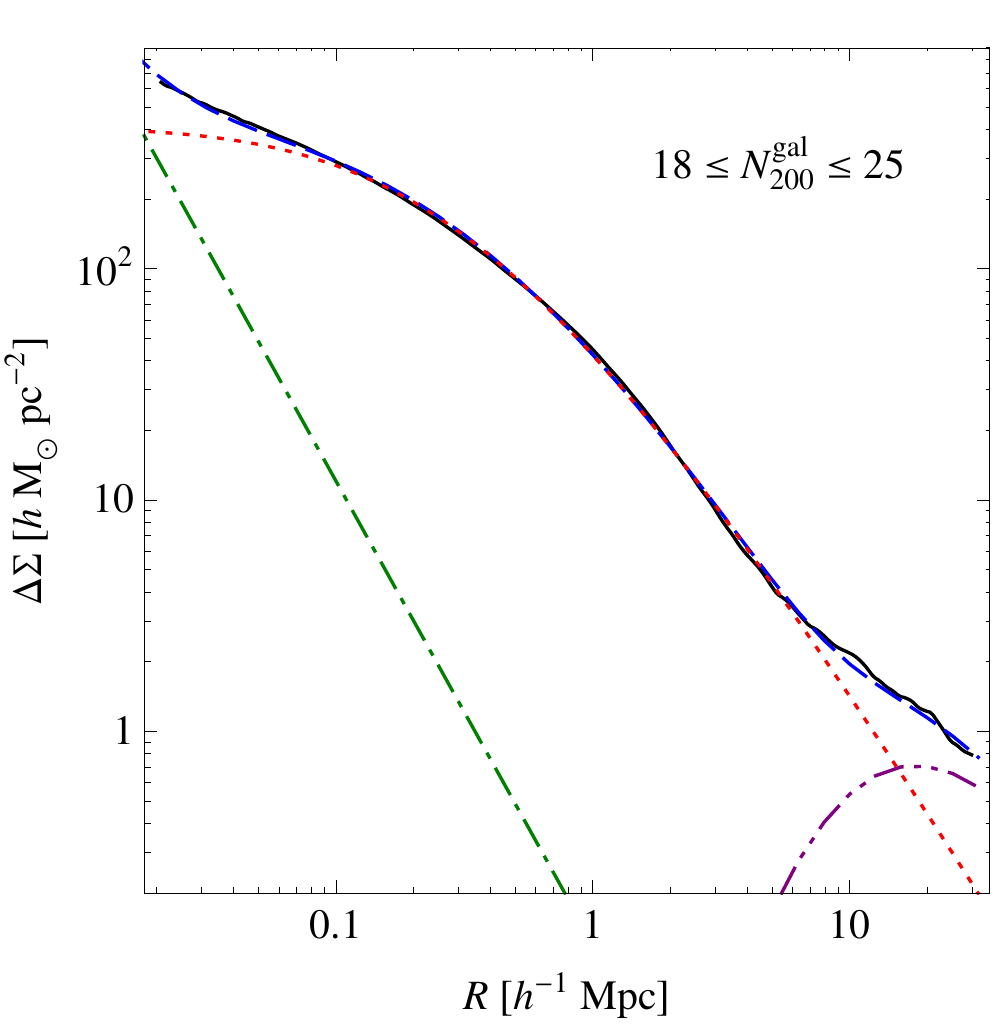}}
\caption{
\label{fig:mass_profile_fit}
Fit to the weak-lensing mass profile $\Delta \Sigma(R)$ as a function of radius $R$ of correctly centred clusters in the MS model. Shown are the measured profiles (solid line), the 3-component fit (dashed line), the central galaxy contribution (dash dotted line), the DM halo contribution (dotted line), and the contribution from neighbouring masses (dash-dot-dotted line).
}
\end{figure}
%================================================

%================================================
\begin{figure}
\centerline{\includegraphics[width=0.8\linewidth]{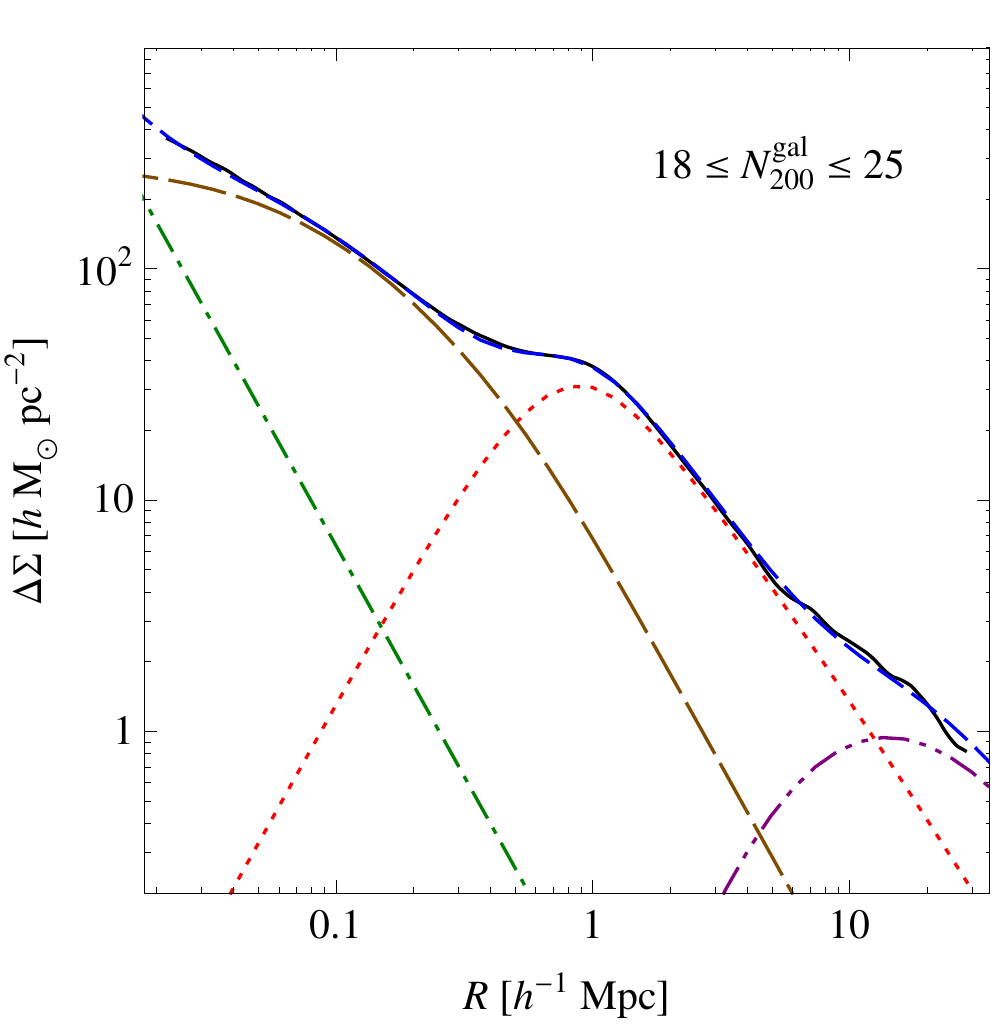}}
\caption{
\label{fig:mass_profile_fit_off_center}
Fit to the weak-lensing mass profile $\Delta \Sigma(R)$ as a function of radius $R$ of incorrectly centred clusters in the MS model. Shown are the measured profiles (solid line), the 4-component fit (short-dashed line), the central galaxy contribution (dash-dotted line), the DM halo contribution (dotted line), the contribution from neighbouring masses (dash-dot-dotted line), and the subhalo contribution (long-dashed line).
}
\end{figure}
%================================================

The resulting fit to the average weak-lensing mass profile of correctly centred clusters is shown in Fig.~\ref{fig:mass_profile_fit} for clusters in the MS model with richness $18\leq \N200 \leq 25$. A fit of similar quality can also be obtained for the other richness bins.

We now turn to the contribution from incorrectly centred clusters. Figure~\ref{fig:center_vs_off_center_mass_profiles} shows that the average profile of the incorrectly centred clusters increases with decreasing radius even for small radii. This is a consequence of the particular choice for the apparent centre of these clusters in our simulation, namely another massive cluster galaxy within a massive subhalo. If the apparent cluster centres were chosen randomly, one would expect the profile to decrease with decreasing radius for $R<1h^{-1}\,\Mpc$ \citep{JohnstonEtal2007_SDSS_cluster_wl_II_arXiv}.

To obtain a good fit to the average mass profiles of incorrectly centred clusters, we thus need four components: a central galaxy component (again modelled as point mass), a mean dark matter halo modelled as an NFW profile convolved with a 2D Gaussian, a contribution from neighbouring masses, and a subhalo, which we model as a truncated NFW profile \citep[][]{BaltzMarshallOguri2009}. As Fig.~\ref{fig:mass_profile_fit_off_center} illustrates, the subhalo component is essential for a good fit to the simulated mass profiles.

%%%%%%%%%%%%%%%%%%%%%%%%%%%%%%%%%%%%%%%%%%%%%%%%%
\section{Summary and Discussion}
\label{sec:discussion}
%%%%%%%%%%%%%%%%%%%%%%%%%%%%%%%%%%%%%%%%%%%%%%%%%

In this work, we have used $N$-body simulations of cosmic structure formation with semi-analytic galaxy formation modelling to test this modelling and to investigate how the properties of optically selected galaxy groups and clusters depend on cosmological parameters. We have created catalogues of simulated galaxy groups/clusters from model galaxy catalogues \citep[by][]{DeLuciaBlaizot2007,WangEtal2008}. We have computed weak-lensing mass profiles and various other properties for these clusters as a function of cluster richness $\N200$ and luminosity $\L200$ \citep[as defined by the maxBCG algorithm of][]{KoesterEtal2007_MaxBCG}, and compared the results to observations of clusters in the SDSS \citep[][]{SheldonEtal2009_SDSS_cluster_wl_I,JohnstonEtal2007_SDSS_cluster_wl_II_arXiv,ReyesEtal2008}.

We find that the simulated weak-lensing mass profiles and the observed profiles of the SDSS maxBCG clusters agree remarkably well in detailed shape and amplitude. Moreover, all simulations reproduce the observed mass-richness relation within $\sim30\%$ over the whole range probed by the SDSS clusters. The MS and the WMAP1-A simulation also yield cluster abundances very similar to the observed abundances. This shows that the models considered here provide a good description of the masses, density profiles and optical properties of galaxy clusters as well as the optical cluster selection and richness estimation of the maxBCG algorithm. Evidently, using mock galaxy catalogues based on a large high-resolution $\Lambda$CDM structure formation simulation and semi-analytic galaxy formation models makes it possible to create very realistic mock cluster catalogues for surveys like the SDSS (when problems with the ridgeline colours are overcome).

Although the underlying $N$-body simulations assume different cosmological parameters that lead to different DM halo abundances \citep[][]{SpringelEtal2005_Millennium, WangEtal2008}, all the galaxy models used here are able to reproduce the observed abundance and two-point correlations of galaxies reasonably well through careful adjustment of their star-formation efficiency and feedback parameters \citep[][]{DeLuciaBlaizot2007,WangEtal2008}. They differ, however, in their predicted cluster abundance as a function of $\N200$. The MS model and the WMAP1-A model, both of which use cosmological parameters based on 1st-year WMAP data \citep[][]{SpergelEtal2003_WMAP_1stYear_Data}, predict cluster abundances that are compatible with the observed values \citep[][]{SheldonEtal2009_SDSS_cluster_wl_I}, whereas the WMAP3-A and WMAP3-B models, which use parameters consistent with the WMAP 3rd-year results \citep[][]{SpergelEtal2007_WMAP_3rdYear_Data}, yield abundances that are lower by a factor 2-3.

The cluster masses predicted as a function of richness $\N200$ or luminosity $\L200$ also differ for the different galaxy formation models. At given richness or luminosity, the MS and WMAP1-A models produce clusters that are up to 30\% more massive than observed, while the WMAP3 cluster masses are similar to or lower than the observed masses.

The different abundances and average cluster masses in our various galaxy formation models are primarily a reflection of the different underlying cosmologies. Because halos in the WMAP3 cosmology are less massive than in the MS/WMAP1 cosmology, the WMAP3 models need more efficient star formation to match the observed galaxy number densities. This produces more ridgeline galaxies in a dark matter halo of given mass. Consequently, clusters at a given richness are less massive in the WMAP3 models than in the MS and WMAP1-A models. Nevertheless, the higher star formation efficiency in the WMAP3 models does not fully compensate for the lower number of massive halos. So there are fewer rich clusters in the WMAP3 models than in the MS and WMAP1-A models.

The lower cluster masses in the observations compared to the MS model suggests that our Universe would be better described by $\sigma_8<0.9$. The higher observed rich cluster abundance than in the WMAP3 models suggests that $\sigma_8>0.72$. Thus, more precise agreement between predicted and observed cluster properties is expected for an intermediate value $0.72<\sigma_8<0.9$. This corroborates the findings by \citet[][]{RozoEtal2009_MaxBCG_V_cosmology_arXiv} that the SDSS maxBCG cluster data favour $\sigma_8\approx0.83$.

Our results confirm that the mass distribution of clusters of given richness is well described by a log-normal distribution. This justifies both the assumption of such distributions and the specific scatter values adopted in previous work which modelled stacked cluster mass profiles or reconstructed cluster mass functions. Fits to the stacked mass profile of clusters whose centre has erroneously been identified with a non-central cluster galaxy should take into account a halo component associated with this non-central galaxy in addition to its stellar mass, the halo component of the main cluster, and surrounding large-scale structure.

Our simulations required many simplifying assumptions about the richness measurements (e.g. about limiting magnitudes, colours, or projection effects), which could result in biased estimates. Although predictions for cluster abundances or for cluster masses are individually subject to such modelling errors, they do not vary independently for varying assumptions about the
richness measurement. For example, assuming a fainter magnitude limit results in lower cluster masses \emph{and} higher cluster numbers. Thus, it proves difficult to decrease average cluster masses for MS model in order to better match the observations without producing too many rich clusters. Similarly, the cluster numbers in the WMAP3 models can only be brought into agreement with the observed counts at the cost of cluster masses which are too small.

Similar reasoning reveals that changes to the galaxy formation description alone, though they could change the number, colour, or brightness of the model galaxies, could only lead to better agreement for either cluster abundance or cluster mass, but not both. This stems from the constraint that the abundances and masses of clusters as a function of richness must `add up' to reproduce the underlying cluster mass function regardless of the specific galaxy formation model. To reach better agreement for both, the number of massive clusters has to be adjusted, too, by changing the cosmology.

Our results demonstrate that, on the one hand, the cluster mass-richness relation and the cluster abundance-richness function together provide strong constraints for the cosmology even without perfect knowledge of the galaxy formation physics. On the other hand, our findings show that the cluster abundance and cluster mass as functions of richness can also be used directly, in addition to galaxy abundance and galaxy two-point correlation, to test the galaxy models.

In future work, one should test how much the agreement between observed and predicted cluster properties can be improved by choosing different cosmological parameters for the simulations \citep[e.g. the values currently favoured by other observations such as][]{KomatsuEtal2009}. In addition, the galaxy models discussed here should be improved to better match the observed colours of ridgeline galaxies at higher redshift. (If we had not adjusted the ridgeline colour selection by hand, the models would have contained almost no clusters with $z>0.25$.) Future simulations should also probe other cosmologies and galaxy models. These simulations need to have a much larger volume than the WMAP simulations to have good enough statistics to match up-coming observations. (The statistical errors on the mass-richness relation in the WMAP simulations are comparable to the uncertainties in current observations.)

More realistic modelling of cluster selection and characterisation could be achieved by running the observational cluster-finding algorithms on mock galaxy catalogues created from the simulations. Moreover, ray-tracing techniques could be used to simulate realistically the weak-lensing mass measurements and to assess their statistical accuracy and possible systematic uncertainties.

Finally we note that there is some tension between our finding that the abundances and weak-lensing masses of clusters favour cosmologies with a normalisation $\sigma_8\approx0.8$ over those with $\sigma_8=0.72$ \citep[in agreement with estimates by][]{RozoEtal2009_MaxBCG_V_cosmology_arXiv,KomatsuEtal2009} and the findings by \citet{LiEtal2009} and \citet{CacciatoEtal2009} that galaxy-galaxy lensing and galaxy clustering data are consistent with $\sigma_8=0.73$. It is beyond the scope of this paper to analyse possible reasons for this discrepancy and whether it can be resolved with better handling of measurement systematics or improved structure formation models, but this should be done in future work.

%%%%%%%%%%%%%%%%%%%%%%%%%%%%%%%%%%%%%%%%%%%%%%%%%%%%%%%%%%%%%%%%%%%%%%%
\section*{Acknowledgments}
We thank Gabriella De Lucia, J{\'er}{\'e}my Blaizot, Jie Wang, Ben Koester, Erin Sheldon, Jan Hartlap, and Peter Schneider for helpful discussions. We thank Jie Wang and collaborators for granting access to their simulation data. We thank Erin Sheldon for providing the SDSS cluster mass profiles. This work was supported by the DFG within the Priority Programme 1177 under the projects SCHN 342/6 and WH 6/3.

%%%%%%%%%%%%%%%%%%%%%%%%%%%%%%%%%%%%%%%%%%%%%%%%%%%%%%%%%%%%%%%%%%%%%%%
%\bibliographystyle{mnbst}
%\bibliography{/users/shilbert/Documents/bibliography/Astro}
%\bibliography{C:/Users/Stefan/Documents/Arbeit/bibliography/Astro}

%%%%%%%%%%%%%%%%%%%%%%%%%%%%%%%%%%%%%%%%%%%%%%%%%

%%%%%%%%%%%%%%%%%%%%%%%%%%%%%%%%%%%%%%%%%%%%%%%%%

%%%%%%%%%%%%%%%%%%%%%%%%%%%%%%%%%%%%%%%%%%%%%%%%%%%%%%%%%%%%%%%%%%%%%%%

%================================================
\begin{table*}
\center
  \caption{
\label{tab:clusters_summary_MS}
The average properties of the model clusters in the MS as inferred from the galaxy model by \citet{DeLuciaBlaizot2007} in bins of richness $\N200$:
the comoving number density $n$, the average number $\ev{\Nint}$ of ridgeline galaxies, the average number $\ev{\N200}$ of ridgeline galaxies within $\Rgal$, the average cluster radius $\ev{\Rgal}$, the average ridgeline $i$-band luminosity $\ev{\L200}$ within $\Rgal$, the average cluster masses $\ev{\Mcrit}$, and the standard deviation $\sigma_{\log_{10}(\Mcrit)}$ of the logarithmic cluster mass. For bins containing $<7$ independent clusters, $\sigma_{\log_{10}(\Mcrit)}$ is omitted.
}
\begin{tabular}{r c c r r c c c c}
\hline
\hline
bin   & $\N200$ & $n$ [$h^{3}\Mpc^{-3}$]& $\ev{\Nint}$  & $\ev{\N200}$ & $\ev{\Rgal}$ [$h^{-1} \Mpc$] & $\ev{\L200}$ [$h^{-1} \Lsolar$] & $\ev{\Mcrit}$ [$h^{-1} \Msolar$]  & $\sigma_{\log_{10}(\Mcrit)}$    \\
\hline
  1 & 3       & $1.2\times 10^{-4}$ &    2.7 &    3.0 &   0.37 & $5.9\times 10^{10}$ & $1.1\times 10^{13}$ &   0.32 \\
  2 & 4       & $5.6\times 10^{-5}$ &    3.6 &    4.0 &   0.45 & $8.0\times 10^{10}$ & $1.6\times 10^{13}$ &   0.31 \\
  3 & 5       & $3.0\times 10^{-5}$ &    4.6 &    5.0 &   0.52 & $1.0\times 10^{11}$ & $2.3\times 10^{13}$ &   0.28 \\
  4 & 6       & $1.8\times 10^{-5}$ &    5.7 &    6.0 &   0.58 & $1.2\times 10^{11}$ & $2.9\times 10^{13}$ &   0.26 \\
  5 & 7       & $1.2\times 10^{-5}$ &    6.7 &    7.0 &   0.64 & $1.5\times 10^{11}$ & $3.6\times 10^{13}$ &   0.24 \\
  6 & 8       & $8.3\times 10^{-6}$ &    7.7 &    8.0 &   0.69 & $1.7\times 10^{11}$ & $4.3\times 10^{13}$ &   0.23 \\
  7 & 9-11    & $1.4\times 10^{-5}$ &    9.5 &    9.8 &   0.78 & $2.1\times 10^{11}$ & $5.5\times 10^{13}$ &   0.21 \\
  8 & 12-17   & $1.1\times 10^{-5}$ &   13.7 &   14.0 &   0.95 & $2.9\times 10^{11}$ & $8.4\times 10^{13}$ &   0.19 \\
  9 & 18-25   & $5.4\times 10^{-6}$ &   20.7 &   20.9 &   1.18 & $4.3\times 10^{11}$ & $1.3\times 10^{14}$ &   0.17 \\
 10 & 26-40   & $3.0\times 10^{-6}$ &   30.6 &   31.2 &   1.44 & $6.4\times 10^{11}$ & $2.0\times 10^{14}$ &   0.15 \\
 11 & 41-70   & $1.3\times 10^{-6}$ &   48.9 &   51.0 &   1.81 & $1.0\times 10^{12}$ & $3.4\times 10^{14}$ &   0.14 \\
 12 & 71-220  & $3.3\times 10^{-7}$ &   86.6 &   93.8 &   2.36 & $1.9\times 10^{12}$ & $6.6\times 10^{14}$ &   0.15 \\
 13 & 221-660 & $1.1\times 10^{-8}$ &  235.5 &  269.1 &   3.69 & $5.1\times 10^{12}$ & $2.1\times 10^{15}$ &    --  \\
\hline
\end{tabular}
\end{table*}
%================================================

%================================================
\begin{table*}
\center
  \caption{
\label{tab:clusters_summary_WMAP1_A}
The properties of the cluster sample in the WMAP1-A model in bins of richness $\N200$ (see Table~\ref{tab:clusters_summary_MS} for a description of the listed quantities).
}
\begin{tabular}{r c c r r c c c c}
\hline
\hline
bin   & $\N200$ & $n$ [$h^{3}\Mpc^{-3}$]& $\ev{\Nint}$  & $\ev{\N200}$ & $\ev{\Rgal}$ [$h^{-1} \Mpc$] & $\ev{\L200}$ [$h^{-1} \Lsolar$] & $\ev{\Mcrit}$ [$h^{-1} \Msolar$]  & $\sigma_{\log_{10}(\Mcrit)}$    \\
\hline
  1 & 3       & $1.3\times 10^{-4}$ &    2.7 &    3.0 &   0.37 & $6.0\times 10^{10}$ & $1.0\times 10^{13}$ &   0.32 \\
  2 & 4       & $6.2\times 10^{-5}$ &    3.6 &    4.0 &   0.45 & $8.1\times 10^{10}$ & $1.5\times 10^{13}$ &   0.30 \\
  3 & 5       & $3.4\times 10^{-5}$ &    4.7 &    5.0 &   0.52 & $1.0\times 10^{11}$ & $2.1\times 10^{13}$ &   0.26 \\
  4 & 6       & $2.2\times 10^{-5}$ &    5.8 &    6.0 &   0.58 & $1.2\times 10^{11}$ & $2.7\times 10^{13}$ &   0.24 \\
  5 & 7       & $1.4\times 10^{-5}$ &    6.8 &    7.0 &   0.64 & $1.4\times 10^{11}$ & $3.2\times 10^{13}$ &   0.22 \\
  6 & 8       & $9.9\times 10^{-6}$ &    7.8 &    8.0 &   0.69 & $1.7\times 10^{11}$ & $3.9\times 10^{13}$ &   0.22 \\
  7 & 9-11    & $1.7\times 10^{-5}$ &    9.4 &    9.8 &   0.77 & $2.1\times 10^{11}$ & $5.2\times 10^{13}$ &   0.22 \\
  8 & 12-17   & $1.1\times 10^{-5}$ &   13.5 &   13.8 &   0.94 & $2.9\times 10^{11}$ & $7.3\times 10^{13}$ &   0.19 \\
  9 & 18-25   & $5.8\times 10^{-6}$ &   22.1 &   21.3 &   1.20 & $4.3\times 10^{11}$ & $1.2\times 10^{14}$ &   0.16 \\
 10 & 26-40   & $3.6\times 10^{-6}$ &   30.9 &   31.8 &   1.46 & $6.6\times 10^{11}$ & $1.8\times 10^{14}$ &   0.16 \\
 11 & 41-70   & $7.9\times 10^{-7}$ &   40.2 &   44.5 &   1.71 & $8.9\times 10^{11}$ & $2.3\times 10^{14}$ &    --  \\
 12 & 71-220  & $5.1\times 10^{-7}$ &   94.4 &  108.1 &   2.49 & $2.1\times 10^{12}$ & $1.0\times 10^{15}$ &    --  \\
\hline
\end{tabular}
\end{table*}
%================================================

%================================================
\begin{table*}
\center
  \caption{
\label{tab:clusters_summary_WMAP3_B}
The properties of the cluster sample in the WMAP3-B model in bins of richness $\N200$ (see Table~\ref{tab:clusters_summary_MS} for a description of the listed quantities).
}
\begin{tabular}{r c c r r c c c c}
\hline
\hline
bin   & $\N200$ & $n$ [$h^{3}\Mpc^{-3}$]& $\ev{\Nint}$  & $\ev{\N200}$ & $\ev{\Rgal}$ [$h^{-1} \Mpc$] & $\ev{\L200}$ [$h^{-1} \Lsolar$] & $\ev{\Mcrit}$ [$h^{-1} \Msolar$]  & $\sigma_{\log_{10}(\Mcrit)}$    \\
\hline
  1 & 3       & $9.6\times 10^{-5}$ &    2.7 &    3.0 &   0.37 & $5.9\times 10^{10}$ & $8.8\times 10^{12}$ &   0.33 \\
  2 & 4       & $4.2\times 10^{-5}$ &    3.6 &    4.0 &   0.44 & $8.2\times 10^{10}$ & $1.4\times 10^{13}$ &   0.33 \\
  3 & 5       & $2.0\times 10^{-5}$ &    4.6 &    5.0 &   0.51 & $1.0\times 10^{11}$ & $1.9\times 10^{13}$ &   0.29 \\
  4 & 6       & $1.1\times 10^{-5}$ &    5.7 &    6.0 &   0.58 & $1.3\times 10^{11}$ & $2.4\times 10^{13}$ &   0.27 \\
  5 & 7       & $6.2\times 10^{-6}$ &    6.9 &    7.0 &   0.64 & $1.5\times 10^{11}$ & $3.0\times 10^{13}$ &   0.26 \\
  6 & 8       & $3.9\times 10^{-6}$ &    7.9 &    8.0 &   0.70 & $1.7\times 10^{11}$ & $3.5\times 10^{13}$ &   0.24 \\
  7 & 9-11    & $6.3\times 10^{-6}$ &    9.8 &    9.8 &   0.79 & $2.2\times 10^{11}$ & $4.8\times 10^{13}$ &   0.25 \\
  8 & 12-17   & $4.7\times 10^{-6}$ &   14.6 &   13.7 &   0.94 & $3.0\times 10^{11}$ & $7.2\times 10^{13}$ &   0.20 \\
  9 & 18-25   & $2.2\times 10^{-6}$ &   21.5 &   21.0 &   1.18 & $4.6\times 10^{11}$ & $1.2\times 10^{14}$ &    --  \\
 10 & 26-40   & $6.6\times 10^{-7}$ &   33.8 &   29.2 &   1.36 & $6.3\times 10^{11}$ & $1.7\times 10^{14}$ &    --  \\
 11 & 41-70   & $3.0\times 10^{-7}$ &   64.4 &   52.9 &   1.74 & $1.1\times 10^{12}$ & $2.6\times 10^{14}$ &    --  \\
 12 & 71-220  & $1.6\times 10^{-7}$ &   72.6 &   80.4 &   2.03 & $1.5\times 10^{12}$ & $3.8\times 10^{14}$ &    --  \\
\hline
\end{tabular}
\end{table*}
%================================================

%================================================
\begin{table*}
\center
  \caption{
\label{tab:clusters_summary_WMAP3_C}
The properties of the cluster sample in the WMAP3-C model in bins of richness $\N200$ (see Table~\ref{tab:clusters_summary_MS} for a description of the listed quantities).
}
\begin{tabular}{r c c r r c c c c}
\hline
\hline
bin   & $\N200$ & $n$ [$h^{3}\Mpc^{-3}$]& $\ev{\Nint}$  & $\ev{\N200}$ & $\ev{\Rgal}$ [$h^{-1} \Mpc$] & $\ev{\L200}$ [$h^{-1} \Lsolar$] & $\ev{\Mcrit}$ [$h^{-1} \Msolar$]  & $\sigma_{\log_{10}(\Mcrit)}$    \\
\hline
  1 & 3       & $1.1\times 10^{-4}$ &    2.6 &    3.0 &   0.37 & $5.8\times 10^{10}$ & $7.8\times 10^{12}$ &   0.33 \\
  2 & 4       & $4.9\times 10^{-5}$ &    3.6 &    4.0 &   0.44 & $7.9\times 10^{10}$ & $1.2\times 10^{13}$ &   0.32 \\
  3 & 5       & $2.5\times 10^{-5}$ &    4.6 &    5.0 &   0.52 & $1.0\times 10^{11}$ & $1.6\times 10^{13}$ &   0.28 \\
  4 & 6       & $1.5\times 10^{-5}$ &    5.6 &    6.0 &   0.58 & $1.2\times 10^{11}$ & $2.0\times 10^{13}$ &   0.26 \\
  5 & 7       & $9.1\times 10^{-6}$ &    6.5 &    7.0 &   0.63 & $1.4\times 10^{11}$ & $2.5\times 10^{13}$ &   0.24 \\
  6 & 8       & $5.6\times 10^{-6}$ &    7.3 &    8.0 &   0.68 & $1.7\times 10^{11}$ & $2.9\times 10^{13}$ &   0.24 \\
  7 & 9-11    & $7.9\times 10^{-6}$ &    9.3 &    9.7 &   0.77 & $2.1\times 10^{11}$ & $3.9\times 10^{13}$ &   0.24 \\
  8 & 12-17   & $6.6\times 10^{-6}$ &   14.6 &   14.2 &   0.97 & $3.1\times 10^{11}$ & $6.8\times 10^{13}$ &   0.19 \\
  9 & 18-25   & $2.6\times 10^{-6}$ &   20.4 &   20.5 &   1.16 & $4.3\times 10^{11}$ & $9.5\times 10^{13}$ &   0.20 \\
 10 & 26-40   & $1.2\times 10^{-6}$ &   30.1 &   29.6 &   1.41 & $6.5\times 10^{11}$ & $1.5\times 10^{14}$ &    --  \\
 11 & 41-70   & $2.5\times 10^{-7}$ &   65.7 &   53.2 &   1.77 & $1.1\times 10^{12}$ & $2.5\times 10^{14}$ &    --  \\
 12 & 71-220  & $2.7\times 10^{-7}$ &   77.5 &   82.0 &   2.09 & $1.6\times 10^{12}$ & $3.7\times 10^{14}$ &    --  \\
\hline
\end{tabular}
\end{table*}
%================================================

%================================================
\begin{table*}
\center
  \caption{
\label{tab:clusters_summary_MS_L}
The average properties of the model clusters in the MS model in bins of luminosity $\L200$ (see Table~\ref{tab:clusters_summary_MS} for a description of the listed quantities).
}
\begin{tabular}{r c c r r c c c c}
\hline
\hline
bin   & $\L200$ [$10^{10}h^{-2}\Lsolar$] & $n$ [$h^{3}\Mpc^{-3}$]& $\ev{\Nint}$ & $\ev{\N200}$ & $\ev{\Rgal}$ [$h^{-1} \Mpc$] & $\ev{\L200}$ [$h^{-1} \Lsolar$] & $\ev{\Mcrit}$ [$h^{-1} \Msolar$]  & $\sigma_{\log_{10}(\Mcrit)}$    \\
\hline
  1 &   5.00 -   6.24 & $4.1\times 10^{-5}$ &    2.9 &    3.3 &   0.39 & $5.6\times 10^{10}$ & $1.0\times 10^{13}$ &   0.27 \\
  2 &   5.24 -   7.80 & $4.4\times 10^{-5}$ &    3.3 &    3.6 &   0.42 & $7.0\times 10^{10}$ & $1.3\times 10^{13}$ &   0.27 \\
  3 &   7.80 -   9.74 & $3.9\times 10^{-5}$ &    3.8 &    4.1 &   0.46 & $8.7\times 10^{10}$ & $1.8\times 10^{13}$ &   0.27 \\
  4 &   9.74 -  12.2  & $3.1\times 10^{-5}$ &    4.6 &    4.9 &   0.51 & $1.1\times 10^{11}$ & $2.4\times 10^{13}$ &   0.26 \\
  5 &  12.2  -  15.2  & $2.2\times 10^{-5}$ &    5.7 &    5.9 &   0.58 & $1.4\times 10^{11}$ & $3.2\times 10^{13}$ &   0.24 \\
  6 &  15.2  -  19.0  & $1.6\times 10^{-5}$ &    7.2 &    7.4 &   0.66 & $1.7\times 10^{11}$ & $4.2\times 10^{13}$ &   0.21 \\
  7 &  19.0  -  23.7  & $1.1\times 10^{-5}$ &    9.2 &    9.3 &   0.76 & $2.1\times 10^{11}$ & $5.6\times 10^{13}$ &   0.19 \\
  8 &  27.7  -  29.6  & $7.7\times 10^{-6}$ &   11.7 &   11.8 &   0.87 & $2.6\times 10^{11}$ & $7.4\times 10^{13}$ &   0.18 \\
  9 &  29.6  -  36.9  & $5.2\times 10^{-6}$ &   14.9 &   15.0 &   0.99 & $3.3\times 10^{11}$ & $9.5\times 10^{13}$ &   0.16 \\
 10 &  36.9  -  46.1  & $3.7\times 10^{-6}$ &   18.7 &   18.9 &   1.12 & $4.1\times 10^{11}$ & $1.2\times 10^{14}$ &   0.16 \\
 11 &  46.1  -  57.6  & $2.5\times 10^{-6}$ &   23.7 &   23.8 &   1.26 & $5.1\times 10^{11}$ & $1.6\times 10^{14}$ &   0.15 \\
 12 &  57.6  -  71.9  & $1.6\times 10^{-6}$ &   29.3 &   29.8 &   1.40 & $6.4\times 10^{11}$ & $2.0\times 10^{14}$ &   0.13 \\
 13 &  71.9  -  89.8  & $9.6\times 10^{-7}$ &   37.0 &   37.9 &   1.57 & $8.0\times 10^{11}$ & $2.6\times 10^{14}$ &   0.13 \\
 14 &  89.8  - 112.1  & $6.5\times 10^{-7}$ &   46.7 &   47.5 &   1.76 & $1.0\times 10^{12}$ & $3.3\times 10^{14}$ &   0.12 \\
 15 & 112.1  - 140    & $4.1\times 10^{-7}$ &   56.8 &   59.7 &   1.94 & $1.2\times 10^{12}$ & $4.1\times 10^{14}$ &   0.13 \\
 16 & 140    - 450    & $3.6\times 10^{-7}$ &   84.7 &   91.2 &   2.31 & $1.9\times 10^{12}$ & $6.4\times 10^{14}$ &   0.16 \\
\hline
\end{tabular}
\end{table*}
%================================================

%================================================
\begin{table*}
\center
  \caption{
\label{tab:clusters_summary_WMAP1_A_L}
The average properties of the model clusters in the WMAP1-A model in bins of luminosity $\L200$ (see Table~\ref{tab:clusters_summary_MS} for a description of the listed quantities).
}
\begin{tabular}{r c c r r c c c c}
\hline
\hline
bin   & $\L200$ [$10^{10}h^{-2}\Lsolar$] & $n$ [$h^{3}\Mpc^{-3}$]& $\ev{\Nint}$ & $\ev{\N200}$ & $\ev{\Rgal}$ [$h^{-1} \Mpc$] & $\ev{\L200}$ [$h^{-1} \Lsolar$] & $\ev{\Mcrit}$ [$h^{-1} \Msolar$]  & $\sigma_{\log_{10}(\Mcrit)}$     \\
\hline
  1 &   5.00 -   6.24 & $4.4\times 10^{-5}$ &    2.9 &    3.3 &   0.39 & $5.6\times 10^{10}$ & $9.8\times 10^{12}$ &   0.27 \\
  2 &   5.24 -   7.80 & $5.0\times 10^{-5}$ &    3.2 &    3.6 &   0.42 & $7.0\times 10^{10}$ & $1.3\times 10^{13}$ &   0.26 \\
  3 &   7.80 -   9.74 & $4.4\times 10^{-5}$ &    3.9 &    4.2 &   0.46 & $8.7\times 10^{10}$ & $1.7\times 10^{13}$ &   0.26 \\
  4 &   9.74 -  12.2  & $3.7\times 10^{-5}$ &    4.7 &    5.0 &   0.52 & $1.1\times 10^{11}$ & $2.2\times 10^{13}$ &   0.24 \\
  5 &  12.2  -  15.2  & $2.5\times 10^{-5}$ &    5.6 &    5.9 &   0.57 & $1.4\times 10^{11}$ & $2.8\times 10^{13}$ &   0.22 \\
  6 &  15.2  -  19.0  & $1.7\times 10^{-5}$ &    7.2 &    7.2 &   0.65 & $1.7\times 10^{11}$ & $3.8\times 10^{13}$ &   0.21 \\
  7 &  19.0  -  23.7  & $1.2\times 10^{-5}$ &    9.1 &    9.2 &   0.75 & $2.1\times 10^{11}$ & $5.0\times 10^{13}$ &   0.19 \\
  8 &  27.7  -  29.6  & $9.3\times 10^{-6}$ &   11.4 &   11.5 &   0.85 & $2.6\times 10^{11}$ & $6.6\times 10^{13}$ &   0.17 \\
  9 &  29.6  -  36.9  & $5.9\times 10^{-6}$ &   14.9 &   14.5 &   0.97 & $3.3\times 10^{11}$ & $8.5\times 10^{13}$ &   0.17 \\
 10 &  36.9  -  46.1  & $4.3\times 10^{-6}$ &   21.0 &   19.8 &   1.15 & $4.1\times 10^{11}$ & $1.1\times 10^{14}$ &   0.15 \\
 11 &  46.1  -  57.6  & $2.5\times 10^{-6}$ &   22.8 &   24.0 &   1.26 & $5.1\times 10^{11}$ & $1.3\times 10^{14}$ &    --  \\
 12 &  57.6  -  71.9  & $1.4\times 10^{-6}$ &   29.2 &   30.4 &   1.42 & $6.5\times 10^{11}$ & $1.8\times 10^{14}$ &    --  \\
 13 &  71.9  -  89.8  & $1.7\times 10^{-6}$ &   36.2 &   37.6 &   1.59 & $8.0\times 10^{11}$ & $2.2\times 10^{14}$ &    --  \\
 14 &  89.8  - 112.1  & $3.9\times 10^{-7}$ &   39.4 &   43.5 &   1.65 & $9.6\times 10^{11}$ & $2.3\times 10^{14}$ &    --  \\
 15 & 112.1  - 140    & $1.8\times 10^{-7}$ &   55.0 &   63.9 &   1.88 & $1.2\times 10^{12}$ & $2.3\times 10^{14}$ &    --  \\
 16 & 140    - 450    & $5.1\times 10^{-7}$ &   94.5 &  108.1 &   2.50 & $2.1\times 10^{12}$ & $1.0\times 10^{15}$ &    --  \\
\hline
\end{tabular}
\end{table*}
%================================================

%================================================
\begin{table*}
\center
  \caption{
\label{tab:clusters_summary_WMAP3_B_L}
The average properties of the model clusters in the WMAP3-B model in bins of luminosity $\L200$ (see Table~\ref{tab:clusters_summary_MS} for a description of the listed quantities).
}
\begin{tabular}{r c c r r c c c c}
\hline
\hline
bin   & $\L200$ [$10^{10}h^{-2}\Lsolar$] & $n$ [$h^{3}\Mpc^{-3}$]& $\ev{\Nint}$ & $\ev{\N200}$ & $\ev{\Rgal}$ [$h^{-1} \Mpc$] & $\ev{\L200}$ [$h^{-1} \Lsolar$] & $\ev{\Mcrit}$ [$h^{-1} \Msolar$]  & $\sigma_{\log_{10}(\Mcrit)}$     \\
\hline
  1 &   5.00 -   6.24 & $3.0\times 10^{-5}$ &    2.8 &    3.3 &   0.39 & $5.6\times 10^{10}$ & $8.5\times 10^{12}$ &   0.29 \\
  2 &   5.24 -   7.80 & $3.2\times 10^{-5}$ &    3.2 &    3.6 &   0.41 & $7.0\times 10^{10}$ & $1.1\times 10^{13}$ &   0.29 \\
  3 &   7.80 -   9.74 & $3.0\times 10^{-5}$ &    3.8 &    4.0 &   0.45 & $8.7\times 10^{10}$ & $1.5\times 10^{13}$ &   0.29 \\
  4 &   9.74 -  12.2  & $2.1\times 10^{-5}$ &    4.4 &    4.6 &   0.49 & $1.1\times 10^{11}$ & $1.9\times 10^{13}$ &   0.27 \\
  5 &  12.2  -  15.2  & $1.4\times 10^{-5}$ &    5.3 &    5.4 &   0.54 & $1.4\times 10^{11}$ & $2.4\times 10^{13}$ &   0.26 \\
  6 &  15.2  -  19.0  & $8.9\times 10^{-6}$ &    6.8 &    6.8 &   0.63 & $1.7\times 10^{11}$ & $3.3\times 10^{13}$ &   0.24 \\
  7 &  19.0  -  23.7  & $5.1\times 10^{-6}$ &    8.9 &    8.7 &   0.74 & $2.1\times 10^{11}$ & $4.2\times 10^{13}$ &   0.22 \\
  8 &  27.7  -  29.6  & $3.7\times 10^{-6}$ &   11.2 &   10.8 &   0.83 & $2.6\times 10^{11}$ & $5.9\times 10^{13}$ &   0.22 \\
  9 &  29.6  -  36.9  & $2.6\times 10^{-6}$ &   14.0 &   13.6 &   0.93 & $3.3\times 10^{11}$ & $7.5\times 10^{13}$ &    --  \\
 10 &  36.9  -  46.1  & $1.8\times 10^{-6}$ &   18.4 &   17.5 &   1.06 & $4.1\times 10^{11}$ & $1.0\times 10^{14}$ &    --  \\
 11 &  46.1  -  57.6  & $9.6\times 10^{-7}$ &   25.5 &   22.2 &   1.22 & $5.1\times 10^{11}$ & $1.4\times 10^{14}$ &    --  \\
 12 &  57.6  -  71.9  & $5.0\times 10^{-7}$ &   30.2 &   26.2 &   1.32 & $6.4\times 10^{11}$ & $1.6\times 10^{14}$ &    --  \\
 13 &  71.9  -  89.8  & $2.6\times 10^{-7}$ &   43.8 &   35.4 &   1.45 & $7.9\times 10^{11}$ & $2.1\times 10^{14}$ &    --  \\
 14 &  89.8  - 112.1  & $1.5\times 10^{-7}$ &   59.7 &   49.1 &   1.68 & $1.0\times 10^{12}$ & $2.3\times 10^{14}$ &    --  \\
 15 & 112.1  - 140    & $1.3\times 10^{-7}$ &   67.9 &   61.4 &   1.85 & $1.2\times 10^{12}$ & $2.9\times 10^{14}$ &    --  \\
 16 & 140    - 450    & $1.4\times 10^{-7}$ &   72.6 &   81.3 &   2.03 & $1.5\times 10^{12}$ & $3.8\times 10^{14}$ &    --  \\
\hline
\end{tabular}
\end{table*}
%================================================

%================================================
\begin{table*}
\center
  \caption{
\label{tab:clusters_summary_WMAP3_C_L}
The average properties of the model clusters in the WMAP3-C model in bins of luminosity $\L200$ (see Table~\ref{tab:clusters_summary_MS} for a description of the listed quantities).
}
\begin{tabular}{r c c r r c c c c}
\hline
\hline
bin   & $\L200$ [$10^{10}h^{-2}\Lsolar$] & $n$ [$h^{3}\Mpc^{-3}$]& $\ev{\Nint}$ & $\ev{\N200}$ & $\ev{\Rgal}$ [$h^{-1} \Mpc$] & $\ev{\L200}$ [$h^{-1} \Lsolar$] & $\ev{\Mcrit}$ [$h^{-1} \Msolar$]  & $\sigma_{\log_{10}(\Mcrit)}$    \\
\hline
  1 &   5.00 -   6.24 & $3.9\times 10^{-5}$ &    2.9 &    3.3 &   0.39 & $5.6\times 10^{10}$ & $8.1\times 10^{12}$ &   0.29 \\
  2 &   5.24 -   7.80 & $3.9\times 10^{-5}$ &    3.3 &    3.6 &   0.42 & $7.0\times 10^{10}$ & $1.0\times 10^{13}$ &   0.30 \\
  3 &   7.80 -   9.74 & $3.3\times 10^{-5}$ &    3.8 &    4.1 &   0.46 & $8.7\times 10^{10}$ & $1.3\times 10^{13}$ &   0.29 \\
  4 &   9.74 -  12.2  & $2.6\times 10^{-5}$ &    4.5 &    4.8 &   0.50 & $1.1\times 10^{11}$ & $1.7\times 10^{13}$ &   0.27 \\
  5 &  12.2  -  15.2  & $1.6\times 10^{-5}$ &    5.5 &    5.9 &   0.56 & $1.4\times 10^{11}$ & $2.2\times 10^{13}$ &   0.24 \\
  6 &  15.2  -  19.0  & $1.0\times 10^{-5}$ &    6.7 &    7.0 &   0.63 & $1.7\times 10^{11}$ & $2.9\times 10^{13}$ &   0.23 \\
  7 &  19.0  -  23.7  & $6.2\times 10^{-6}$ &    8.8 &    8.8 &   0.73 & $2.1\times 10^{11}$ & $3.8\times 10^{13}$ &   0.21 \\
  8 &  27.7  -  29.6  & $4.7\times 10^{-6}$ &   11.6 &   11.7 &   0.86 & $2.6\times 10^{11}$ & $5.3\times 10^{13}$ &   0.20 \\
  9 &  29.6  -  36.9  & $3.5\times 10^{-6}$ &   14.9 &   15.0 &   0.99 & $3.3\times 10^{11}$ & $6.9\times 10^{13}$ &   0.17 \\
 10 &  36.9  -  46.1  & $2.1\times 10^{-6}$ &   18.3 &   17.9 &   1.09 & $4.1\times 10^{11}$ & $9.3\times 10^{13}$ &    --  \\
 11 &  46.1  -  57.6  & $1.1\times 10^{-6}$ &   23.3 &   23.2 &   1.24 & $5.1\times 10^{11}$ & $1.3\times 10^{14}$ &    --  \\
 12 &  57.6  -  71.9  & $6.4\times 10^{-7}$ &   31.2 &   27.9 &   1.38 & $6.4\times 10^{11}$ & $1.5\times 10^{14}$ &    --  \\
 13 &  71.9  -  89.8  & $3.7\times 10^{-7}$ &   37.9 &   33.9 &   1.49 & $7.8\times 10^{11}$ & $1.8\times 10^{14}$ &    --  \\
 14 &  89.8  - 112.1  & $1.4\times 10^{-7}$ &   57.8 &   48.0 &   1.70 & $1.0\times 10^{12}$ & $2.2\times 10^{14}$ &    --  \\
 15 & 112.1  - 140    & $1.4\times 10^{-7}$ &   67.5 &   59.7 &   1.87 & $1.2\times 10^{12}$ & $2.7\times 10^{14}$ &    --  \\
 16 & 140    - 450    & $2.8\times 10^{-7}$ &   77.3 &   81.1 &   2.07 & $1.6\times 10^{12}$ & $3.7\times 10^{14}$ &    --  \\
\hline
\end{tabular}
\end{table*}
%================================================

%%%%%%%%%%%%%%%%%%%%%%%%%%%%%%%%%%%%%%%%%%%%%%%%%%%%%%%%%%%%%%%%%%%%%%%
\end{document}